\pdfoutput=1
\documentclass[aps,prd,twocolumn,superscriptaddress,preprintnumbers,floatfix,nofootinbib]{revtex4-2}

\usepackage{graphicx}
\usepackage{amsmath}
\usepackage{color}
\usepackage{dcolumn}
\usepackage{tensor}
\usepackage{bm}
\usepackage{makecell}
\usepackage[T1]{fontenc}
\usepackage[utf8]{inputenc}
\usepackage{microtype}
\usepackage{etoolbox}
\usepackage{amssymb}
\usepackage{mathrsfs}
\usepackage{accents}
\usepackage[normalem]{ulem}
\usepackage[table,dvipsnames]{xcolor}
\usepackage[colorlinks,urlcolor=NavyBlue,citecolor=NavyBlue,linkcolor=NavyBlue,pdfusetitle]{hyperref}
\usepackage[all]{hypcap}
\usepackage[inline,shortlabels]{enumitem}
\usepackage{braket}

\usepackage{array}
\usepackage{diagbox}
\usepackage{color}
\usepackage{colortbl}
\usepackage{hhline}
\usepackage{multirow}

\graphicspath{{./}{./plots/}}

\def\l@section{\@dottedtocline{1}{2em}{2em}}

\newcommand{\scri}{\mathscr{I}}

\renewcommand{\Re}{\text{Re}}

\newcommand{\figPNBMScomparison}{%
\begin{figure*}
	\label{fig:PNBMScomparison}
	\centering
	\includegraphics[width=\textwidth]{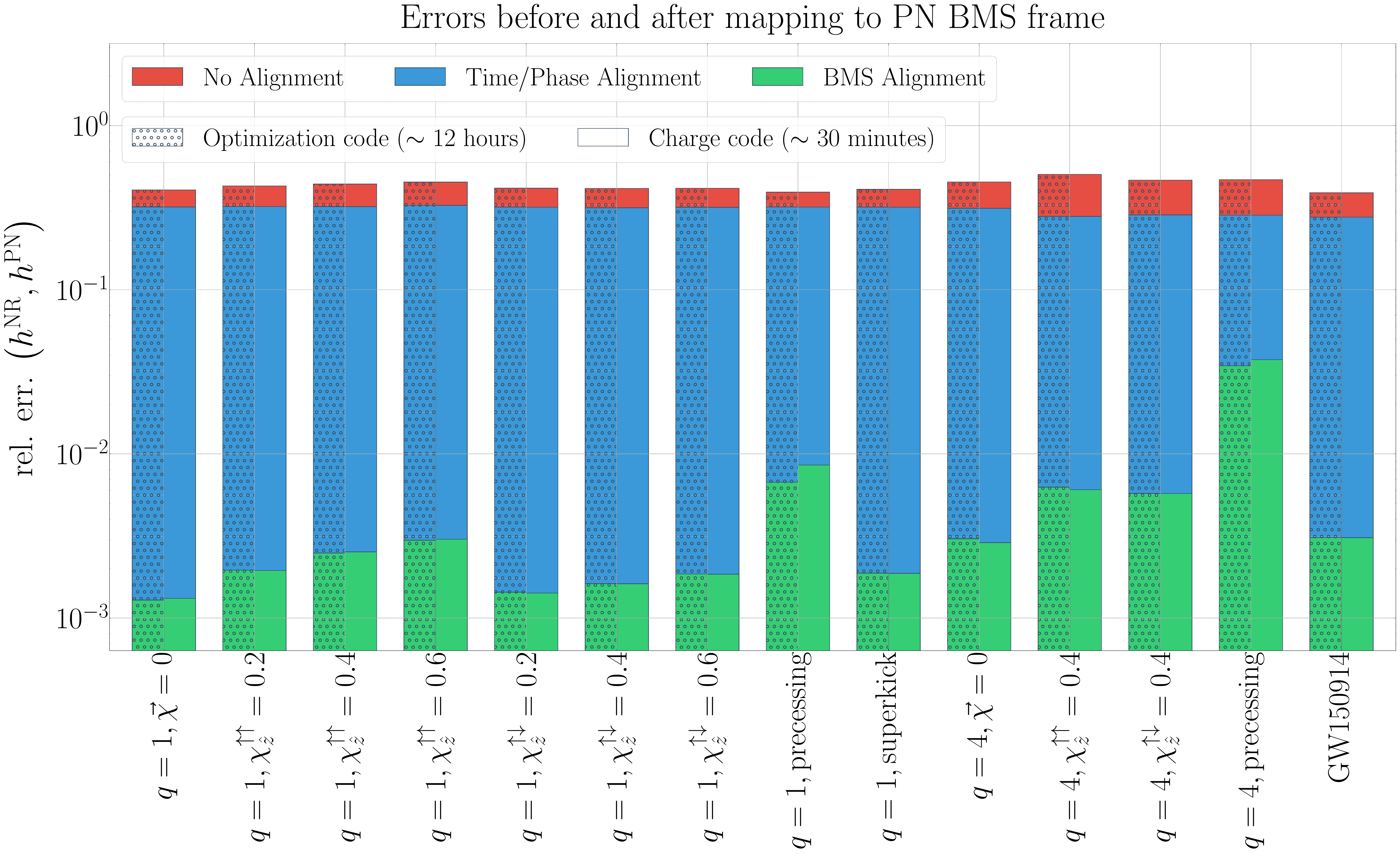}
	\caption{%
		Comparison of the new charge-based frame fixing
		method (plain bars) to the previous method, which determined the BMS transformations by minimizing the $L^{2}$ norm
		of the absolute difference of the NR and PN strains
		(patterned bars) when mapping to the PN BMS frame. The top of
		each bar is the relative error between the NR and PN strain
		waveforms over a three-orbit window that begins $1200M$
		past the initial time of the simulation. The red bars
		correspond to the error when no frame fixing is performed;
		the blue bars correspond to the error when the time and
		phase freedom is fixed; the green bars correspond to the
		error when the whole BMS frame is fixed. These simulations
		correspond to those found in Table~\ref{tab:runs}.}
\end{figure*}%
}

\newcommand{\figsuperrestcomparison}{%
\begin{figure*}
	\label{fig:superrestcomparison}
	\centering
	\includegraphics[width=\textwidth]{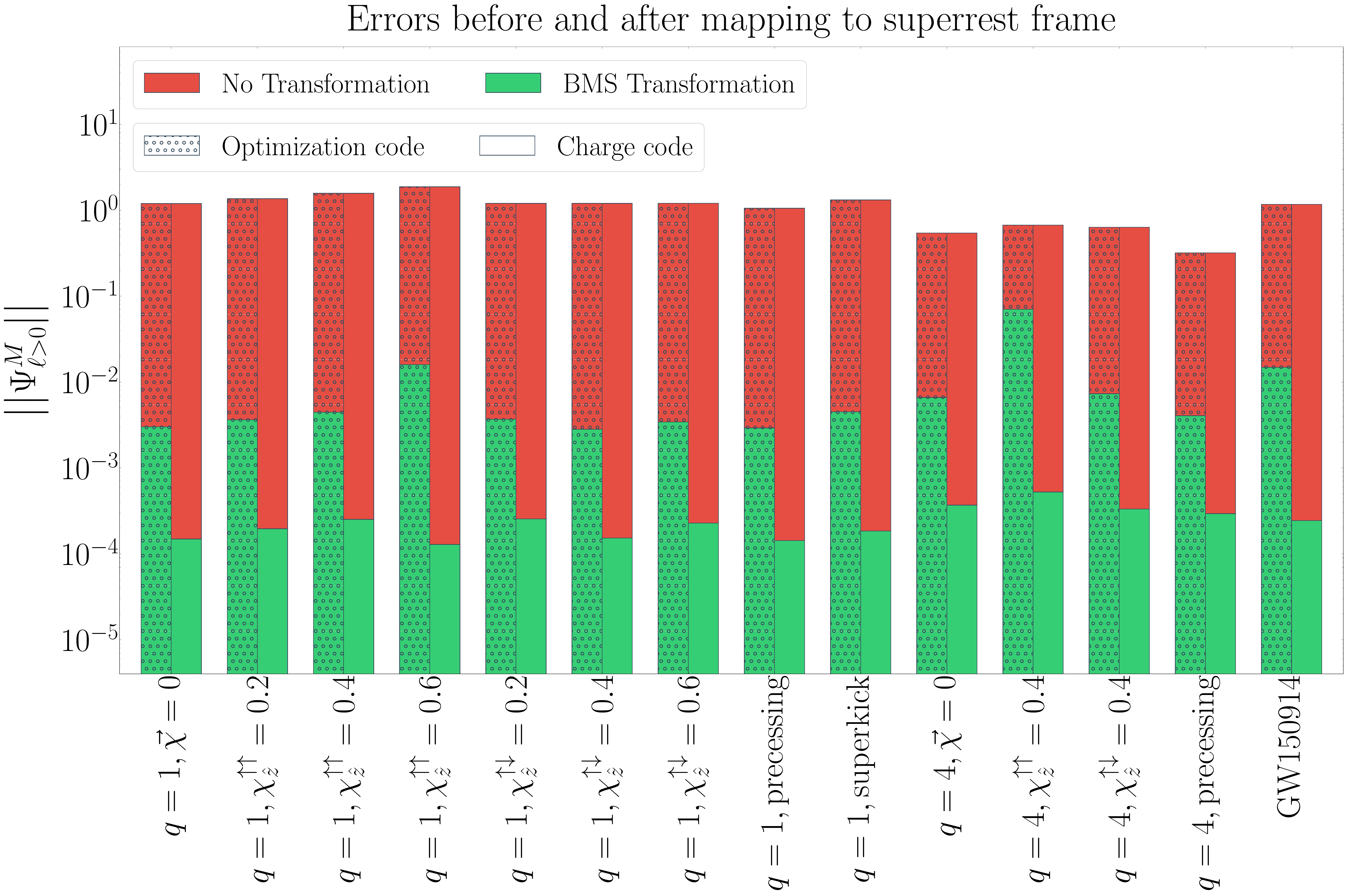}
	\caption{Comparison of the new charge-based frame fixing
		method (plain bars) to the previous method, which
		determined the BMS transformations by minimizing the
		$L^{2}$ norm of the $\ell>0$ components of the Moreschi supermomentum (patterned bars) when mapping to the
		superrest frame. The top of each bar is the norm of the
		$\ell>0$ components of the Moreschi supermomentum computed
		using Eq.~\eqref{eq:Moreschisupermomentum}. The red bars
		correspond to the norm when no frame fixing is performed,
		while the green bars correspond to the norm when the BMS
		frame is fixed. These simulations correspond to those found
		in Table~\ref{tab:runs}.}
\end{figure*}%
}

\newcommand{\figPNcomparison}{%
\begin{figure*}
\label{fig:PNcomparison}
\centering
\includegraphics[width=\textwidth]{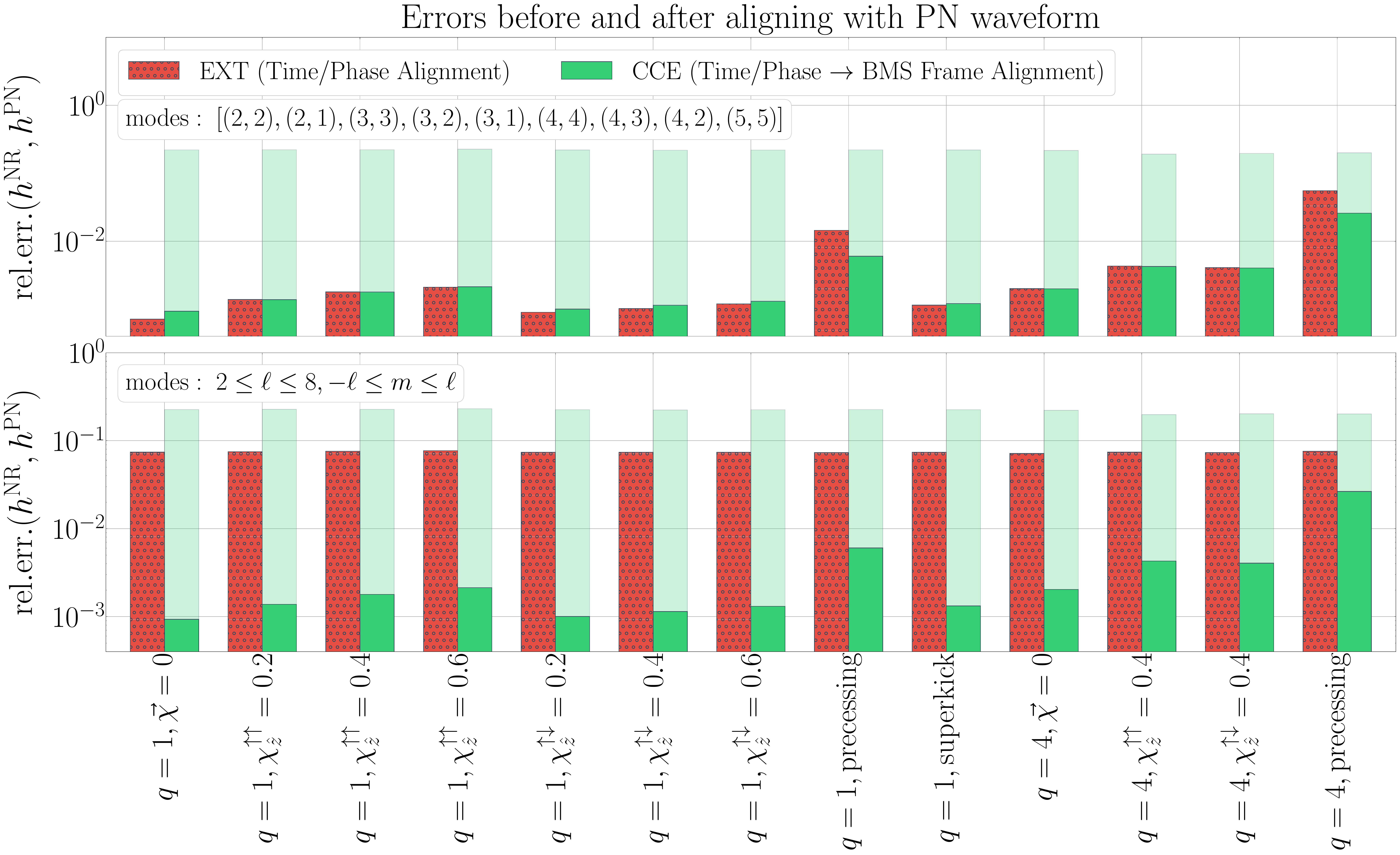}
\caption{Comparison of the relative error between a CCE waveform and
a PN waveform (green) to the relative error between an extrapolated
waveform and a PN waveform (red). The relative error is computed over
a three-orbit window that begins at $1200M$ past the initial time
of the simulation. For the CCE waveforms, we show two errors: one
where the error is computed when the time and phase freedom is
fixed (faint) and one where the error is computed when the BMS
freedom is fixed (full). With the extrapolated waveforms, we only fix
the time and phase freedom. The top panel shows the relative error when only
the $(2,2)$, $(2,1)$, $(3,3)$, $(3,2)$, $(3,1)$, $(4,4)$, $(4,3)$,
$(4,2)$, and $(5,5)$ modes are included, while the bottom panel
shows the error when every mode up to $\ell=8$ is included. These
simulations correspond to those found in Table~\ref{tab:runs}.}
\end{figure*}%
}

\newcommand{\figQNMcomparison}{%
\begin{figure*}
\label{fig:QNMcomparison}
\centering
\includegraphics[width=\textwidth]{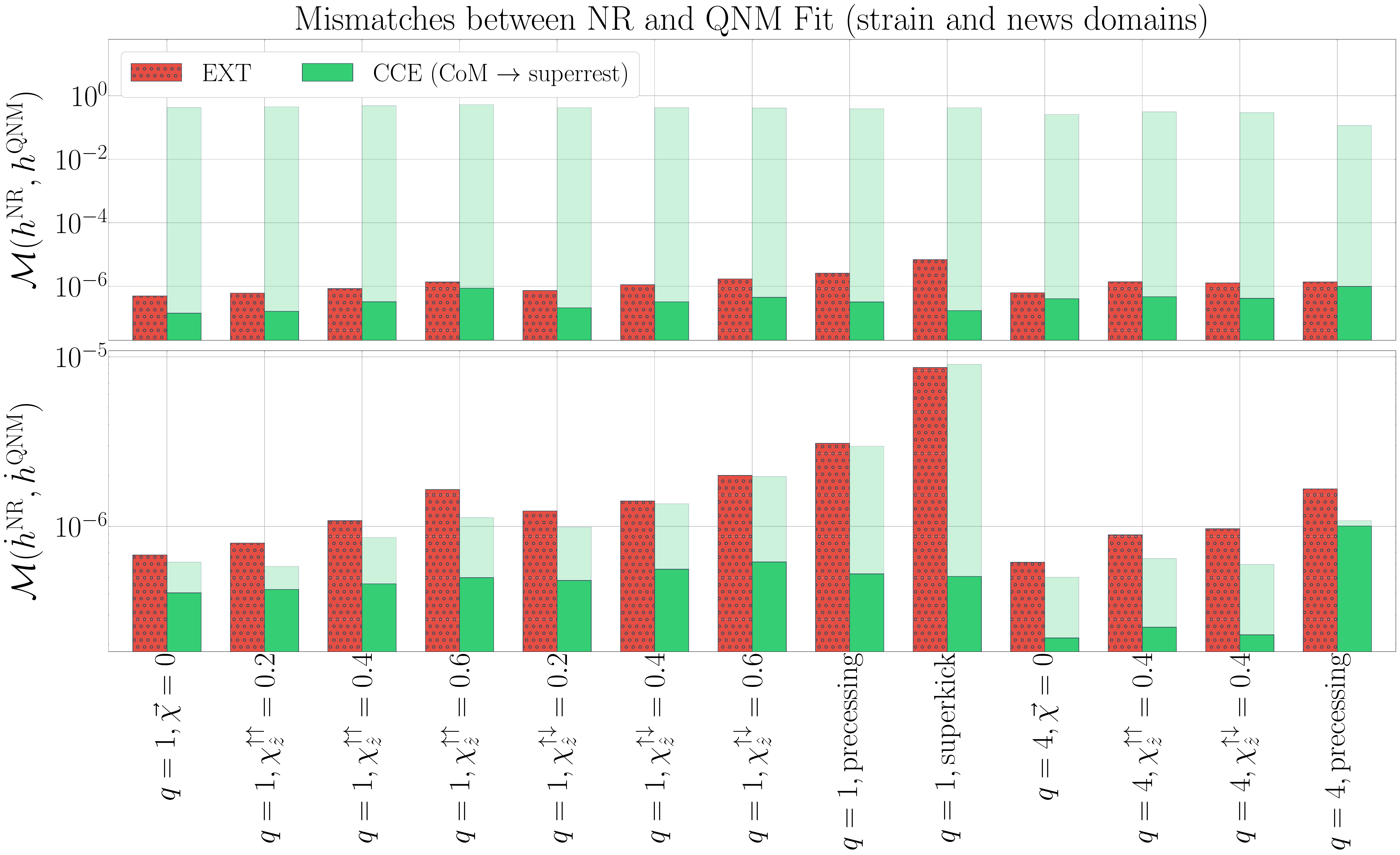}
\caption{Comparison of the mismatch (the top of each bar) between
a CCE waveform and the best-fit QNM model built from 100 modes
chosen using the algorithm presented in~\cite{MaganaZertuche:2021syq}
(green), to the mismatch obtained when using  an extrapolated waveform
instead of the CCE waveform (red). The mismatch is computed from
the peak of the $L^{2}$ norm of the news onward to be consistent
with~\cite{MaganaZertuche:2021syq}. The mass and spin of the
remnant black hole are obtained from the Bondi rest mass and the
spin charge, Eq.~\eqref{eq:spincharge}. For
the CCE waveforms, we show two mismatches: one where the mismatch
is computed using a CCE waveform in the center-of-mass frame of
the remnant black hole (faint) and one where the CCE waveform is
in the superrest frame of the remnant black hole (full). For the
extrapolated waveforms, we only change each mode by a constant so
that the waveform decays to zero at the end of the simulation. The
top panel shows the mismatch between the strain waveforms, while the
bottom panel shows the mismatch between the news waveforms. These
simulations correspond to those found in Table~\ref{tab:runs}.}
\end{figure*}%
}

\usepackage{orcidlink}

\newcommand{\Cornell}{\affiliation{Cornell Center for Astrophysics and Planetary
		Science, Cornell University, Ithaca, New York 14853, USA}}
\newcommand{\Caltech}{\affiliation{Theoretical Astrophysics 350-17, California
		Institute of Technology, Pasadena, CA 91125, USA}}
\newcommand{\MaxPlanck}{\affiliation{Max Planck Institute for Gravitational
		Physics (Albert Einstein Institute), Am M{\"u}hlenberg 1, D-14476 Potsdam,
		Germany}}
\newcommand{\OleMiss}{\affiliation{Department of Physics and Astronomy,
		University of Mississippi, University, Mississippi 38677, USA}}

\begin{document}

\title{Fixing the BMS frame of numerical relativity waveforms with BMS charges}
	
\author{Keefe Mitman
	\orcidlink{0000-0003-0276-3856}}
\email{kmitman@caltech.edu}
\Caltech
\author{Leo C. Stein
	\orcidlink{0000-0001-7559-9597}}
\email{lcstein@olemiss.edu}
\OleMiss
\author{Michael Boyle 
	\orcidlink{0000-0002-5075-5116}}
\Cornell

\author{\\Nils Deppe \orcidlink{0000-0003-4557-4115}} \Caltech
\author{Fran\c{c}ois H\'{e}bert \orcidlink{0000-0001-9009-6955}}\Caltech
\author{Lawrence E.~Kidder \orcidlink{0000-0001-5392-7342}} \Cornell
\author{Jordan Moxon \orcidlink{0000-0001-9891-8677}} \Caltech
\author{Mark A. Scheel
	\orcidlink{0000-0001-6656-9134}}
\Caltech
\author{Saul A. Teukolsky
	\orcidlink{0000-0001-9765-4526}}
\Caltech
\Cornell
\author{William Throwe \orcidlink{0000-0001-5059-4378}} \Cornell
\author{Nils L.~Vu \orcidlink{0000-0002-5767-3949}} \MaxPlanck

\hypersetup{pdfauthor={Mitman et al.}}
	
\date{\today}
	
\begin{abstract}
\noindent The Bondi-van der Burg-Metzner-Sachs (BMS)
group, which uniquely describes the symmetries of
asymptotic infinity and therefore of the gravitational waves
that propagate there, has become increasingly important
for accurate
modeling of waveforms. In particular, waveform models, such as post-Newtonian (PN)
expressions, numerical relativity (NR), and black
hole perturbation theory, produce results that are
in different BMS frames.
Consequently, to build a model
for the waveforms produced during the merging of compact
objects, which ideally would be a hybridization of PN, NR,
and black hole perturbation theory,
one needs a fast and robust method for fixing the BMS
freedoms. In this work, we present the first means of fixing
the entire BMS freedom of NR waveforms to match the frame
of either PN waveforms or black hole perturbation theory.
We achieve this by finding
the BMS transformations that change certain charges
in a prescribed way---e.g., finding the center-of-mass
transformation that maps the center-of-mass charge to a mean of zero. We find that this new method is 20 times
faster, and more correct when mapping to the superrest
frame, than previous methods that relied on optimization
algorithms. Furthermore, in the course of
developing this charge-based
frame fixing method, we compute the PN expression for the
Moreschi supermomentum
to 3PN order without spins and 2PN order
with spins.
This Moreschi supermomentum is effectively equivalent
to the energy flux or the null memory contribution at future
null infinity $\scri^{+}$.
From this PN calculation, we also compute oscillatory ($m\not=0$ modes) and
spin-dependent memory terms that have not been identified previously or have been missing from strain
expressions in the post-Newtonian literature.
\end{abstract}
	
\maketitle
	
\section{Introduction}
\label{sec:introduction}

In the coming years, as gravitational wave detectors such
as LIGO, Virgo, and KAGRA commence their next observing run, the catalog
of astrophysical binary events is predicted to considerably increase~\cite{LIGOScientific:2020ibl,LIGOScientific:2021djp,KAGRA:2013rdx}.
With so many more gravitational-wave events, tests of
Einstein's theory of general relativity can then become even more
robust and informative~\cite{Berti:2018cxi,Berti:2018vdi}. From an
analysis point of view, however, regardless of how much the gravitational-wave transient catalog increases in size over time,
our ability to examine these events will always be limited by the
accuracy of our waveform models, since they serve as the basis against
which we can compare our observations.

Currently, and likely for the foreseeable future, the most accurate
models for the gravitational waves emitted by the most commonly
observed astrophysical binary system, binary black holes (BBHs),
are the waveforms produced by numerical relativity (NR).
Numerical waveforms uniquely maintain their precision
throughout the whole evolution of the binary system. By contrast,
other waveform models can only be considered correct during
certain phases of the binary merger---e.g., post-Newtonian (PN)
theory during inspiral or black hole perturbation theory
during ringdown. However, numerical simulations are finite
in time and thus can never produce the full waveform emitted by a
binary event. Therefore, the best waveform model that we can hope
to build is a hybridization of these models: a waveform that is
PN from far into the past until we approach the merger regime,
numerical throughout the merger phase, and then
black hole perturbation theory during the
final stages of ringdown. But, to construct such a waveform requires
knowledge of how these models are related to each other.
More specifically,
to perform such a hybridization one needs to be able to ensure that
these waveform models are in the same \emph{frame}.

Like any observable in nature, the gravitational waves that we
detect are emitted by a system that is in a certain
frame relative to us. Often, this frame is best
interpreted by using
the symmetries of the system
to understand the transformation between the system and the detector.
For the gravitational radiation that we observe, which for
practical purposes can be interpreted as existing at asymptotic
infinity, the symmetry group is not the usual Poincar\'e group,
but an extension: the \emph{BMS group}~\cite{Bondi,Sachs}.

The BMS group is the symmetry group of asymptotically flat, Lorentzian spacetimes at
null infinity $\scri$~\cite{Bondi,Sachs}. It can be viewed as a combination of the Lorentz
group with an infinite dimensional group of transformations called
\emph{supertranslations}~\cite{Bondi,Sachs}.\footnote{Formally,
the BMS group is understood to be a semidirect product of the
Lorentz group with the infinite-dimensional Abelian group of
supertranslations containing the spacetime translations as a normal
subgroup.} While surprising at first, these supertranslations have a natural origin. Consider, for example, a network of observers
positioned on a sphere of finite radius encompassing a source. With
some effort, these observers could combine their received signals with some understanding of their clocks' synchronization. But,
if we move these observers to asymptotic infinity, then such
a synchronization becomes impossible because of the infinite
separation of the observers. More specifically, they will no longer be in causal contact.
Thus, we can freely time
translate---i.e., supertranslate---each observer without changing
the observable physics. Put differently, supertranslations are time
translations applied to each point on the two-sphere at
asymptotic infinity.  That is, they are simply direction-dependent time
translations. Consequently, supertranslations change the retarded
or Bondi time $u\equiv t-r$ via
\begin{align}
\label{eq:BondiTime}
u'(u,\theta,\phi)&=u-\alpha(\theta,\phi)\nonumber\\
&=u-\sum\limits_{\ell=0}^{\infty}\sum\limits_{|m|\leq\ell}\alpha_{(\ell, m)}Y_{(\ell, m)}(\theta,\phi).
\end{align}
In Eq.~\eqref{eq:BondiTime}, $\alpha(\theta,\phi)$
is the supertranslation parameter,
with $\alpha_{\ell,
m}=(-1)^{m}\overline{\alpha}_{\ell,-m}$ to make sure that the
transformed Bondi time $u'$ is real. Using
spherical harmonics to express supertranslations in components is useful because
we see that the usual
spacetime translations are nothing more than the $\ell\leq1$ supertranslations,
while supertranslations with $\ell\geq2$ are the new, proper
supertranslations~\cite{Boyle:2015nqa,Mitman:2021xkq}.

When we compare
waveform models, we need to make sure that the
Poincar\'e freedoms are equivalently fixed, e.g., to ensure that both models
represent a binary in the center-of-mass frame.
However, because of these new symmetries that arise at asymptotic
infinity through the $\ell\geq2$ supertranslations,
we must also require that the supertranslation freedom of each
model is fixed in an equivalent manner. In the work of~\cite{Mitman:2021xkq}
this task of fixing the BMS frame of a numerical waveform to match
that of a PN system was pursued for the first time.
Reference~\cite{Mitman:2021xkq} did this by minimizing the error between a NR strain waveform and a PN strain waveform, iterating
through various BMS transformations to find the best
transformations to apply to the NR system. Because the BMS group
is infinite dimensional, to restrict the parameter space,
Ref.~\cite{Mitman:2021xkq} restricted the transformations
to only include supertranslations up to $\ell\leq4$,
producing a 22-dimensional parameter space (10 Poincar\'e +
12 supertranslations).\footnote{Actually, their parameter space
was only 16-dimensional because they fixed the center-of-mass
transformation using the system's center-of-mass charge. However, the
following argument that we make is nonetheless valid because of the
large number of parameters that were involved in the minimization.} Because of this, the optimization
algorithm required a large amount of CPU time: nearly 12 hours
per system.\footnote{Moreover,
because the BMS transformations
were found using an optimizer rather than a physically motivated scheme,
the waveforms produced tended to be in an error-minimizing
but physically incorrect BMS frame. See, for example, Fig. 9
of~\cite{Mitman:2021xkq}.} Furthermore, because of this need to
restrict the number of transformations that can be applied for the
sake of the optimization algorithm, it is clear that this method of
fixing the BMS frame will not be sufficient when we want to include
higher order modes in our models, which will be important for future detectors.

In this work, we present a more sophisticated means of fixing the
BMS frame, which relies solely on BMS charges and thus removes the
need for optimization algorithms.\footnote{Note that here we use the term `charge' in a more relaxed way, i.e., not in the sense of a symmetry algebra acting on a system.}
This idea is motivated by the
success that the authors of~\cite{Mitman:2021xkq} had when fixing the
center-of-mass frame with the center-of-mass charge.\footnote{See,
for example, Figs. 4 and 5 of~\cite{Mitman:2021xkq}.} In this
work, the charges used to fix the Poincar\'e frame are the
center-of-mass charge and rotation charge, while the charge used
to constrain the supertranslation freedom is a charge known as the
Moreschi supermomentum, which is an extension of the usual Bondi
four-momentum~\cite{Moreschi:1988pc}. With these charges we then map the
NR waveforms to two unique frames: the PN BMS frame, i.e., the
frame that PN waveforms are in, and the superrest frame at timelike infinity $i^{+}$,
i.e., the frame that quasi-normal modes (QNMs) are computed in. To
map to the PN BMS frame, we find that we need to compute the PN
Moreschi supermomentum to 3PN order with spins
and 2PN order without spins. Upon doing so, we discover
oscillatory memory terms and spin-dependent memory terms that have not been identified or have been missing from the PN strain in the literature.
We find that by fixing the
frame of the numerical waveforms using this charge-based method,
rather than the optimization used in~\cite{Mitman:2021xkq},
we can not only obtain the same, if not better, errors between
waveform models, but also do so in roughly 30 minutes compared to
the previous run time of 12 hours.

\subsection{Overview}
\label{sec:overview}

We organize our computations and results as follows. In
Sec.~\ref{sec:bmstransformations} we
describe how the BMS
transformations transform the
asymptotic variables. We also present the three BMS charges that we will be
using in our analyses: the center-of-mass charge, the
rotation charge, and the Moreschi supermomentum. Following this,
in Sec.~\ref{sec:pnsupermomentum}, we calculate the PN Moreschi supermomentum and present the new PN memory
terms that have been missing from earlier calculations of the
PN strain. Finally, in Sec.~\ref{sec:numerics} we show our
numerical results. In Sec.~\ref{sec:bmsframefixing} we outline our method for fixing the BMS frame with BMS charges. In
Sec.~\ref{sec:comparisonprevresults} we compare this new charge-based
method's ability to map the NR system to either the PN BMS frame or the remnant
black hole's superrest frame to that of the previous method. In
Sec.~\ref{sec:comparison} we then highlight how waveforms
produced via Cauchy-characteristic extraction
are much more applicable and correct than the previously used
extrapolated waveforms, once their BMS frame has been fixed with this new charge-based method.
Lastly, in Appendices~\ref{sec:appendixa}
and~\ref{sec:appendixb} we present the complete results of our PN
calculations for the Moreschi supermomentum and the memory terms
missing from the PN strain.

\subsection{Conventions}
\label{sec:conventions}

We set $c=G=1$ and take $\eta_{\mu\nu}$ to be the $(-,+,+,+)$ Minkowski metric. When working with complex dyads, following the work of~\cite{Iozzo:2021vnq,Mitman:2021xkq}, we use
\begin{align}
\label{eq:dyads}
q_{A}=-\frac{1}{\sqrt{2}}(1,i\sin\theta)\text{ and }q^{A}=-\frac{1}{\sqrt{2}}(1,i\csc\theta),
\end{align}
and write the round metric on the two-sphere $S^{2}$ as $q_{AB}$. The complex dyad obeys the following properties
\begin{align}
q_{A}q^{A}=0,\quad q_{A}\bar{q}^{A}=1,\quad q_{AB}=q_{A}\bar{q}_{B}+\bar{q}_{A}q_{B}.
\end{align}
Note that this convention differs from the related works of~\cite{Mitman:2020pbt,Mitman:2020bjf,Moxon:2021gbv}, which in contrast do not include the $1/\sqrt{2}$ normalization factor on the dyads in Eq.~\eqref{eq:dyads}. We choose this convention because it makes our expressions for the asymptotic charges in Eq.~\eqref{eq:Poincarécharges} more uniform. Nonetheless, for transparency we provide the conversion between our quantities and those of these previous works in Eq.~\eqref{eq:conversion}.

We build spin-weighted fields with the dyads as follows. For a tensor field $W_{A\cdots D}$, the function
\begin{align}
W=W_{A\cdots BC\cdots D}q^{A}\cdots q^{B}\bar{q}^{C}\cdots\bar{q}^{D}
\end{align}
with $m$ factors of $q$ and $n$ factors of $\bar{q}$ has a spin-weight of $s=m-n$. When raising and lowering spin-weights we use the Geroch-Held-Penrose differential spin-weight operators $\eth$ and $\bar{\eth}$~\cite{Geroch:1973am},
\begin{subequations}
	\begin{align}
	\eth W&=(D_{E}W_{A\cdots BC\cdots D})q^{A}\cdots q^{B}\bar{q}^{C}\cdots\bar{q}^{D}q^{E},\\
	\bar{\eth}W&=(D_{E}W_{A\cdots BC\cdots D})q^{A}\cdots q^{B}\bar{q}^{C}\dots\bar{q}^{D}\bar{q}^{E}.
	\end{align}
\end{subequations}
Here, $D_{A}$ is the covariant derivative on the two-sphere. The $\eth$ and $\bar{\eth}$ operators in spherical coordinates are then
\begin{subequations}
	\begin{align}
	\eth W(\theta,\phi)&=-\frac{1}{\sqrt{2}}(\sin\theta)^{+s}(\partial_{\theta}+i\csc\theta\partial_{\phi})\nonumber\\
	&\phantom{=.-\frac{1}{\sqrt{2}}(\sin(\theta))^{s}}\left[(\sin\theta)^{-s}W(\theta,\phi)\right],\\
	\bar{\eth} W(\theta,\phi)&=-\frac{1}{\sqrt{2}}(\sin\theta)^{-s}(\partial_{\theta}-i\csc\theta\partial_{\phi})\nonumber\\
	&\phantom{=.-\frac{1}{\sqrt{2}}(\sin(\theta))^{s}}\left[(\sin\theta)^{+s}W(\theta,\phi)\right].
	\end{align}
\end{subequations}
Thus, when acting on spin-weighted spherical harmonics, these operators produce
\begin{subequations}
	\begin{align}
	\eth(\phantom{}_{s}Y_{(\ell, m)})&=+\frac{1}{\sqrt{2}}\sqrt{(\ell-s)(\ell+s+1)}_{s+1}Y_{(\ell, m)},\\
	\bar{\eth}(\phantom{}_{s}Y_{(\ell, m)})&=-\frac{1}{\sqrt{2}}\sqrt{(\ell+s)(\ell-s+1)}_{s-1}Y_{(\ell, m)}.
	\end{align}
\end{subequations}
We denote the gravitational wave strain\footnote{We explicitly define the strain as described in Appendix C of~\cite{Boyle:2019kee}.} by $h$, which we represent in a spin-weight $-2$ spherical harmonic basis,
\begin{align}
h(u,\theta,\phi)=\sum\limits_{\ell,m}h_{(\ell, m)}(u)\,{}_{-2}Y_{(\ell, m)}(\theta,\phi),
\end{align}
where, again, $u\equiv t-r$ is the Bondi time. We denote
the Weyl scalars by $\Psi_\text{0 -- 4}$. The conversion from the
convention of~\cite{Mitman:2020pbt,Mitman:2020pbt,Moxon:2021gbv}
($\texttt{NR}$\footnote{$\texttt{NR}$ because this
is the convention that corresponds to the outputs of the SXS
simulations.}) to ours ($\texttt{MB}$\footnote{$\texttt{MB}$
because this corresponds to the Moreschi-Boyle convention used in the
works~\cite{Moreschi:1988pc,Moreschi:1998mw,Dain:2000lij,Boyle:2015nqa,Iozzo:2021vnq,Mitman:2021xkq}
and the code
\texttt{scri}~\cite{scri_url,Boyle:2013nka,Boyle:2014ioa,Boyle:2015nqa}.})
is
\begin{align}
\label{eq:conversion}
h^{\texttt{NR}}=2\bar{\sigma}^{\texttt{MB}}\quad\text{and}\quad\Psi_{i}^{\texttt{NR}}=\frac{1}{2}(-\sqrt{2})^{i}\Psi_{i}^{\texttt{MB}}.
\end{align}
Note that we will omit these superscripts and henceforth assume that everything is in the $\texttt{MB}$ convention. 

\section{BMS transformations and charges}
\label{sec:bmstransformations}

As discussed in the introduction, the symmetry group of
asymptotic infinity is not the usual Poincar\'e group, but
the BMS group, in which the spacetime translations are extended through
an infinite-dimensional group of transformations called
supertranslations~\cite{Bondi,Sachs}. Therefore, to understand
the frame of asymptotic radiation, we must understand how the
asymptotic variables transform under
an arbitrary BMS transformation.
Every BMS transformation can be uniquely decomposed as a pure supertranslation
followed by a Lorentz transformation.  In terms of retarded time $u$
and a complex stereographic coordinate $\zeta$ on the two-sphere,
\begin{align}
\left(u,\zeta\right)\equiv\left(t-r,e^{i\phi}\cot\left(\theta/2\right)\right)
\end{align}
a BMS transformation acts on the coordinates as~\cite{Boyle:2015nqa,Moreschi:1998mw}
\begin{align}
\label{eq:coordtransformation}
\left(u,\zeta\right)\rightarrow\left(u',\zeta'\right)=\left(k\left(u-\alpha\right),\frac{a\zeta+b}{c\zeta+d}\right),
\end{align}
where the conformal factor is
\begin{align}
\label{eq:conformalfactor}
k(\zeta,\bar{\zeta})&\equiv\frac{1+\zeta\overline{\zeta}}{\left(a\zeta+b\right)\left(\overline{a}\overline{\zeta}+\overline{b}\right)+\left(c\zeta+d\right)\left(\overline{c}\overline{\zeta}+\overline{d}\right)},
\end{align}
$(a,b,c,d)$ are
complex coefficients satisfying $ad-bc=1$, and the parameter
$\alpha(\zeta,\bar{\zeta})$ is a real-valued
and smooth function on the celestial two-sphere.
The parameters $(a,b,c,d)$ encode Lorentz transformations---both boost and rotations---whereas the function
$\alpha(\zeta,\bar{\zeta})$ describes
supertranslations, and thus also translations.
By examining
how the associated tetrad transforms under a BMS transformation~\cite{Boyle:2015nqa,Moreschi1986OnAM},
one then finds that the shear and the Weyl scalars transform as
\begin{subequations}
\label{eq:vartransformation}
\begin{align}
\label{eq:sheartransformation}
\sigma'&=\frac{e^{2i\lambda}}{k}\left[\sigma-\eth^{2}\alpha\right],\\
\label{eq:psitransformation}
\Psi_{A}'&=\frac{e^{\left(2-A\right)i\lambda}}{k^{3}}\sum\limits_{a=A}^{4}\begin{pmatrix}4-A\\a-A\end{pmatrix}\left(-\frac{\eth u'}{k}\right)^{a-A}\Psi_{a},
\end{align}
\end{subequations}
where $A\in\{0,1,2,3,4\}$ and $\lambda$ is the spin phase~\cite{Boyle:2015nqa,Moreschi1986OnAM}:
\begin{align}
\exp(i\lambda)=\left[\frac{\partial\overline{\zeta'}}{\partial\overline{\zeta}}\left(\frac{\partial\zeta'}{\partial\zeta}\right)^{-1}\right]^{1/2}=\frac{c\zeta+d}{\overline{c}\overline{\zeta}+\overline{d}}.
\end{align}
For a Lorentz transformation parameterized
by $(a,b,c,d)$ and a supertranslation parameterized
by $\alpha$, Eqs.~\eqref{eq:coordtransformation}
and~\eqref{eq:vartransformation} are the primary ingredients
for understanding how the asymptotic variables, namely the
shear as well as the Weyl scalars, transform under a
BMS transformation. All that remains is a method for
finding the necessary frame-fixing BMS transformation, which we outline in
Secs.~\ref{sec:bmstransformationsPoincare} for the Poincar\'e
transformations and~\ref{sec:bmstransformationsupertranslations}
for the proper supertranslations.

\subsection{Poincar\'e charges}
\label{sec:bmstransformationsPoincare}

In this section, following the work
of~\cite{Mitman:2021xkq}, we outline the
Poincar\'e charges that will be used to completely fix the
Poincar\'e transformation freedom of the
asymptotic shear and Weyl scalars. As was shown
in~\cite{Bondi,Barnich:2009se,Strominger:2013jfa,He:2014laa,Kapec:2014opa,Pasterski:2015tva,Barnich:2010eb,Strominger:2014pwa,Flanagan:2015pxa,Mitman:2021xkq},
by examining the expansion of the Bondi-Sachs metric near future null
infinity, one can identify certain functions that
yield the Poincar\'e charges when integrated over the celestial two-sphere.
They are the Bondi
mass aspect $m$, the Lorentz aspect $N$, and the energy moment
aspect $E$, which in the \texttt{MB} convention are
\begin{subequations}
\begin{align}
m(u,\theta,\phi)&\equiv-\text{Re}\left[\Psi_{2}+\sigma\dot{\bar{\sigma}}\right],\\
N(u,\theta,\phi)&\equiv-\left(\Psi_{1}+\sigma\eth\bar{\sigma}+u\eth m+\frac{1}{2}\eth\left(\sigma\bar{\sigma}\right)\right),\\
E(u,\theta,\phi)&\equiv N+u\eth m\nonumber\\
&=-\left(\Psi_{1}+\sigma\eth\bar{\sigma}+\frac{1}{2}\eth\left(\sigma\bar{\sigma}\right)\right).
\end{align}
\end{subequations}
Thus, by defining a collection of spin-0 scalar functions, $\mathbf{n}(\theta,\phi)$,
whose components are unique combinations of the
$\ell\leq1$ spherical harmonics so as to represent one of the four
Cartesian coordinates $t$, $x$, $y$, or $z$, i.e.,
\begin{subequations}
	\begin{align}
	n^{t}&=1\nonumber\\
	&=\sqrt{4\pi}Y_{(0,0)},\\
	n^{x}&=\sin\theta\cos\phi,\nonumber\\
	&=\sqrt{\frac{4\pi}{3}}\left[\frac{1}{\sqrt{2}}\left(Y_{(1,-1)}-Y_{(1,+1)}\right)\right],\\
	n^{y}&=\sin\theta\sin\phi\nonumber\\
	&=\sqrt{\frac{4\pi}{3}}\left[\frac{i}{\sqrt{2}}\left(Y_{(1,-1)}+Y_{(1,+1)}\right)\right],\\
	n^{z}&=\cos\theta\nonumber\\
	&=\sqrt{\frac{4\pi}{3}}Y_{(1,0)},
	\end{align}
\end{subequations}
we can then compute each Cartesian component of the translation, rotation, boost, and center-of-mass charges. For $a\in\{t,x,y,z\}$, these four Poincar\'e charges are
\begin{subequations}
\label{eq:Poincarécharges}
\begin{align}
\label{eq:momentumcharge}
P^{a}(u)&=\frac{1}{4\pi}\int_{S^{2}}n^{a} m\,d\Omega,\\
\label{eq:rotationcharge}
J^{a}(u)&=\frac{1}{4\pi}\int_{S^{2}}\text{Re}\left[\left(\bar{\eth}n^{a}\right)\left(-iN\right)\right]\,d\Omega,\\
\label{eq:boostcharge}
K^{a}(u)&=\frac{1}{4\pi}\int_{S^{2}}\text{Re}\left[\left(\bar{\eth}n^{a}\right)N\right]\,d\Omega,\\
\label{eq:CoMcharge}
G^{a}(u)&=\left(K^{a}+uP^{a}\right)/P^{t}\nonumber\\
&=\frac{1}{4\pi}\int_{S^{2}}\text{Re}\left[\left(\bar{\eth}n^{a}\right)\left(N+u\eth m\right)\right]\,d\Omega/P^{t}.
\end{align}
\end{subequations}
Then, by making use of the orthogonality property of spherical
harmonics, we find that the four vector of a
Poincar\'e charge $\Pi$ is simply
\begin{subequations}
\begin{align}
\Pi^{t}&=\frac{1}{\sqrt{4\pi}}\Pi_{(0,0)},\\
\Pi^{x}&=\frac{1}{\sqrt{4\pi}}\frac{1}{\sqrt{6}}\text{Re}\left[\Pi_{(1,-1)}-\Pi_{(1,+1)}\right],\\
\Pi^{y}&=\frac{1}{\sqrt{4\pi}}\frac{1}{\sqrt{6}}\text{Im}\left[\Pi_{(1,-1)}+\Pi_{(1,+1)}\right],\\
\Pi^{z}&=\frac{1}{\sqrt{4\pi}}\frac{1}{\sqrt{3}}\text{Re}\left[\Pi_{(1,0)}\right],
\end{align}
\end{subequations}
where $\Pi_{(\ell,m)}$ is the $(\ell,m)$ mode of the charge $\Pi$ in
the basis of spin-$s$ spherical harmonics ($s=0$ for $m$, and
$s=1$ for $-iN, N$, and $N+u\eth m$).
These charges in Eqs.~\eqref{eq:Poincarécharges} have simple interpretations by analogy to kinematics in Minkowski space. $P^{a}$ is the total linear momentum, $J^{a}$ is the total angular momentum---orbital plus spin,
$K^{a}$ is proportional to the center-of-mass at time $u=0$, and $G^{a}$ is the center-of-mass at time $u$.
For more on these charges in the PN formulation, see~\cite{deAndrade:2000gf} for an analysis using conservative point particle theory,~\cite{Blanchet:2018yqa} for an analysis when radiation is involved, and~\cite{Compere:2019gft} for a connection to the Poincaré and BMS flux-balance laws.

While the Poincar\'e charges in Eq.~\eqref{eq:Poincarécharges}
are the most natural to work with for fixing frames, since we
will be comparing the frame of numerical waveforms to that of PN
waveforms, we need either PN or BH perturbation theory expressions for
these charges.  Notice, however, that
Eqs.~\eqref{eq:Poincarécharges} contain the Weyl scalars $\Psi_{1}$
and $\Psi_{2}$.
Thus we need the PN expressions for these Weyl scalars, which have not
been computed thus far in the PN literature.\footnote{This would be a valuable calculation to carry out in the future.}
Fortunately, PN waveforms are inherently
constructed in the center-of-mass frame. The center-of-mass
charge $G^{a}$, given in Eq.~\eqref{eq:CoMcharge}, can be taken to
have an average of
zero, i.e., an intercept and a slope of zero when fitted to with a
linear function in time, though it is oscillatory at high enough PN
order. For mapping to the superrest frame, we can also map the center-of-mass charge to have an average of zero, seeing as we want our remnant black hole to asymptote to a stationary Kerr black hole. 

The rotation charge,
Eq.~\eqref{eq:rotationcharge}, however, tends to be some nontrivial function
of time. Therefore, for fixing the rotation freedom to match that
of PN waveforms, we require an alternative rotation
charge that is independent of the $\Psi_{1}$ and $\Psi_{2}$
Weyl scalars. In the work of~\cite{Boyle:2013nka} such a rotation charge was built by finding the angular velocity which
keeps the radiative fields as constant as possible in the corotating frame. Following the notation of~\cite{Boyle:2013nka,OShaughnessy:2011pmr}, this vector, with ``$\cdot$'' the usual dot product,
is
\begin{align}
\label{eq:angularvelocityvector}
\vec{\omega}(u)=-\langle\vec{L}\vec{L}\rangle^{-1}\cdot\langle\vec{L}\partial_{t}\rangle,
\end{align}
where we use the $u$-dependent vector and matrix
\begin{subequations}
\begin{align}
\langle\vec{L}\partial_{t}\rangle^{a}&\equiv\sum\limits_{\ell,m,m'}\text{Im}\left[\bar{f}_{(\ell,m')}\langle\ell,m'|L^{a}|\ell,m\rangle\dot{f}_{(\ell,m)}\right],\\
\langle\vec{L}\vec{L}\rangle^{ab}&\equiv\sum\limits_{\ell,m,m'}\bar{f}_{(\ell,m')}\langle\ell,m'|L^{(a}L^{b)}|\ell,m\rangle f_{(\ell,m)},
\end{align}
\end{subequations}
and $f(u,\theta,\phi)$ is some function that corresponds to the
asymptotic radiation, e.g., the shear $\sigma$ or the news
$\dot{\sigma}$.
This data is represented in the basis $|\ell,m\rangle$ of
spin-weighted spherical harmonics, with time-dependent mode weights
$f_{(\ell,m)}$, i.e. $|f\rangle = \sum_{\ell,m}f_{(\ell,m)}|\ell,m\rangle$.
The operator $\vec{L}$ is the infinitesimal
generator of rotations, whose related charge is the total angular
momentum charge $J^{a}$ provided in Eq.~\eqref{eq:rotationcharge}.
Therefore, when fixing the frame of the waveforms to match that of the PN system, i.e., to match the frames at spacelike infinity, we will use the vector in Eq.~\eqref{eq:angularvelocityvector}.

For fixing the rotation at timelike infinity---the case of a system consisting
of a single black hole---we also choose to use
a rotation charge that is not the charge seen in Eq.~\eqref{eq:rotationcharge}. This
is because at timelike infinity, provided that we are in the
center-of-mass frame of the remnant black hole, we do not expect
there to be any orbital angular momentum contributions, but instead
only spin angular momentum contributions. Thus, for fixing the
rotation freedom it is more useful to work with the
asymptotic dimensionless spin vector~\cite{Iozzo:2021vnq}:
\begin{align}
\label{eq:spincharge}
\vec{\chi}(u)=\frac{\gamma}{M_{\text{B}}^{2}}\left(\vec{J}+\vec{v}\times\vec{K}\right)-\frac{\gamma-1}{M_{\text{B}}^{2}}\left(\hat{v}\cdot\vec{J}\right)\hat{v},
\end{align}
where
\begin{align}
\gamma(u)\equiv\sqrt{1-\left|\vec{v}\right|^{2}}^{-1}
\end{align}
is the Lorentz factor, 
\begin{align}
M_{\text{B}}(u)\equiv\sqrt{-\eta_{\mu\nu}P^{\mu}P^{\nu}}
\end{align}
is the Bondi mass,
\begin{align}
\vec{v}(u)\equiv\vec{P}/P^{t}
\end{align}
is the velocity vector, and the vectors $\vec{J}$ and $\vec{K}$
are the angular momentum and boost Poincar\'e charges, i.e.,
Eqs.~\eqref{eq:rotationcharge} and~\eqref{eq:boostcharge}, evaluated
at the vector $\vec{n}=\left(\vec{x},\vec{y},\vec{z}\right)$. With this
charge, we can then fix the rotation freedom by solving for the
rotation that maps this charge to be parallel
to the positive $z$-axis: the standard convention for studying QNMs. This fully fixes the rotation freedom of our system, up to a $U(1)$ transformation that can be thought of as a constant phase change. For mapping to the PN BMS frame, this remaining $U(1)$ freedom is fixed while running a time and phase alignment that minimizes the residual between the NR and PN strain waveforms. For mapping to the superrest frame at timelike infinity, no phase fixing is performed as it is not necessary for analyzing QNMs.

\subsection{$\ell\geq 2$ supertranslation charge}
\label{sec:bmstransformationsupertranslations}

For fixing the supertranslation freedom of a system, because the supertranslations are just extensions of the usual spacetime translations, it is reasonable to ask if there is a clear extension of the Bondi four-momentum~\cite{Bondi}
\begin{align}
\label{eq:Bondifourmomentum}
P^{a}(u)=\frac{1}{4\pi}\int_{S^{2}}n^{a}\text{Re}\left[\Psi_{2}+\sigma\dot{\bar{\sigma}}\right]\,d\Omega.
\end{align}
As was pointed out by Dray and Streubel~\cite{Dray:1984rfa}, and also independently realized later by Wald and Zoupas~\cite{Wald:1999wa}, the possible choices for a \emph{supermomentum} can be written as
\begin{align}
\label{eq:supermomenta}
\Psi_{p,q}(u,\theta,\phi)=\Psi_{2}+\sigma\dot{\bar{\sigma}}+p\left(\eth^{2}\bar{\sigma}\right)-q\left(\bar{\eth}^{2}\sigma\right),
\end{align}
where $p$ and $q$ are arbitrary real numbers. From
this supermomentum expression, one can show that if $p=q$ then there is
no supermomentum flux in Minkowski space and if $p+q=1$ then
the supermomentum is real~\cite{Dray:1984rfa,Wald:1999wa}. This
leads to a natural choice of supermomentum being the Geroch (G)
supermomentum with $p=q=\frac{1}{2}$, i.e.,
\begin{align}
\Psi_{\text{G}}(u,\theta,\phi)\equiv\Psi_{2}+\sigma\dot{\bar{\sigma}}+\frac{1}{2}\left(\eth^{2}\bar{\sigma}-\bar{\eth}^{2}\sigma\right).
\end{align}
It turns out, though, that in regimes that are nonradiative (i.e., $\dot{\sigma}=0$), the Geroch supermomentum is not changed by a supertranslation since
\begin{align}
\Psi_{\text{G}}'&=\frac{1}{k^{3}}\left(\Psi_{\text{G}}-\frac{1}{2}\left[2\eth\alpha\eth\dot{\bar{\sigma}}+\left(\eth^{2}\alpha\right)\dot{\bar{\sigma}}+(\eth\alpha)^{2}\ddot{\bar{\sigma}}+\text{c.c.}\right]\right)\nonumber\\
&\rightarrow\frac{1}{k^{3}}\Psi_{\text{G}}\quad\left(\text{for }\dot{\sigma}\rightarrow0\right).
\end{align}
Therefore, while the Geroch supermomentum may be an ideal choice for
a physical supermomentum, for fixing the supertranslation freedom of
our system it will instead
be useful to construct a supertranslation charge
that does transform under supertranslations. But what are
the features of the system that we would like to control
using these supertranslation transformations?

When we examine how supertranslations transform
asymptotic radiation,
Eq.~\eqref{eq:sheartransformation} shows that the shear is changed
by a term constant in time and proportional to the supertranslation parameter
$\alpha(\theta,\phi)$. Consequently, because supertranslations affect the value
of the shear even in nonradiative regimes, we can interpret them
as also being related to the gravitational memory effect---i.e.,
the physical observable that corresponds to the permanent net change
in the metric due to the passage of transient gravitational
radiation~\cite{Zeldovich:1974gvh,Christodoulou,Thorne}.\footnote{%
  Supertranslations and the memory effect both correspond to changes
  in the value of the strain: supertranslations are merely gauge
  transformations, while the memory effect can be understood as corresponding to holonomy~\cite{Flanagan:2014kfa,Grant:2021hga,Seraj:2022qqj}.}

Gravitational memory can best be understood
through the supermomentum balance law~\cite{Ashtekar:2019viz},
which says that the real part of the shear can be written as
\begin{align}
\label{eq:reshear}
\text{Re}\left[\bar{\eth}^{2}\sigma\right]&=m+\int_{-\infty}^{u}|\dot{\sigma}|^{2}du+\left(\text{Re}\left[\bar{\eth}^{2}\sigma\right]-m\right)|^{-\infty}\nonumber\\
&=m+\mathcal{E}-M_{\text{ADM}},
\end{align}
where
\begin{align}
\mathcal{E}(u,\theta,\phi)\equiv\int_{-\infty}^{u}|\dot{\sigma}|^{2}du
\end{align}
is proportional to the energy radiated up to time $u$ into the
direction $(\theta,\phi)$, and $M_{\text{ADM}}$ is the Arnowitt-Deser-Misner (ADM)
mass~\cite{ADM}. If one evaluates this equation between
early times and late times (i.e., $\pm\infty$), then the net change
in the shear (the memory) has two unique
contributions: one from the net change in the mass aspect (the
\emph{ordinary}\footnote{Also called linear memory~\cite{Zeldovich:1974gvh}.} memory) and one
from the net change in the system's energy flux (the \emph{null}\footnote{Also
called nonlinear or Christodoulou memory~\cite{Christodoulou,Thorne}} memory). Note that the ADM mass does not contribute to the memory because it is a constant on the two-sphere and therefore has no $\ell\geq2$ components.
Typically, ordinary memory
occurs in systems that have unbound masses, such as hyperbolic black holes,
while null memory occurs in systems that have bound masses,
such as binary black holes.
Furthermore, if one examines how this equation changes
under a BMS transformation, one finds
\begin{subequations}
\label{eq:BondimassEnergytransformations}
\begin{align}
\label{eq:Bondimasstransformation}
m'&=\frac{1}{k^{3}}\left(m+\frac{1}{2}\left[2\eth\alpha\eth\dot{\bar{\sigma}}+\left(\eth^{2}\alpha\right)\dot{\bar{\sigma}}+(\eth\alpha)^{2}\ddot{\bar{\sigma}}+\text{c.c.}\right]\right)\nonumber\\
&\rightarrow\frac{1}{k^{3}}m\quad\left(\text{when nonradiative}\right),\\
\label{eq:Energytransformation}
\mathcal{E}'&=\frac{1}{k^{3}}\left(\mathcal{E}-\eth^{2}\bar{\eth}^{2}\alpha\right)\quad\left(\text{always}\right).
\end{align}
\end{subequations}
So, when written in terms
of its charge ($m$) and flux ($\mathcal{E}$) contributions, the only part of the shear
that transforms under
a supertranslation in nonradiative regimes of $\scri^{+}$
is the energy flux contribution. This suggests that the charge that would be the ideal charge for measuring what supertranslation
we should apply to our system would be something equivalent
to the energy flux. Fortunately, such a charge, the Moreschi
supermomentum, has already been examined by Dain and
Moreschi~\cite{Moreschi1986OnAM,Moreschi:1988pc,Moreschi:1998mw,Dain:2000lij}:
\begin{align}
\label{eq:Moreschisupermomentum}
\Psi_{\text{M}}(u,\theta,\phi)&\equiv\Psi_{2}+\sigma\dot{\bar{\sigma}}+\eth^{2}\bar{\sigma}.
\end{align}
This is simply Eq.~\eqref{eq:supermomenta} with coefficients $p=1$
and $q=0$. By rewriting the first term in
Eq.~\eqref{eq:Moreschisupermomentum} as a time integral,
making use of the Bianchi identity
\begin{align}
\dot{\Psi}_{2}=-\sigma\ddot{\bar{\sigma}}+\eth^{2}\bar{\sigma},
\end{align}
and integrating by parts we find
\begin{align}
\label{eq:psimenergyflux}
\Psi_{\text{M}}&=\left(\int_{-\infty}^{u}\dot{\Psi}_{2}\,du-M_{\text{ADM}}\right)+\sigma\dot{\bar{\sigma}}+\eth^{2}\bar{\sigma}\nonumber\\
&=\left(\int_{-\infty}^{u}\left(-\sigma\ddot{\bar{\sigma}}+\eth^{2}\bar{\sigma}\right)du-M_{\text{ADM}}\right)+\sigma\dot{\bar{\sigma}}+\eth^{2}\bar{\sigma}\nonumber\\
&=\int_{-\infty}^{u}|\dot{\sigma}|^{2}du-M_{\text{ADM}}\nonumber\\
&=\mathcal{E}-M_{\text{ADM}}.
\end{align}
Because of this, the Moreschi supermomentum can also be thought of as the memory part of the mass moment.\footnote{See, for example, Eq. (2.26) of~\cite{Favata:2008yd}.}
Furthermore, by using Eq.~\eqref{eq:Energytransformation} with Eq.~\eqref{eq:psimenergyflux}, one finds that the Moreschi supermomentum transforms as
\begin{align}
\label{eq:psimtransformation}
\Psi_{\text{M}}'=\frac{1}{k^{3}}\left(\Psi_{\text{M}}-\eth^{2}\bar{\eth}^{2}\alpha\right).
\end{align}
Equations~\eqref{eq:reshear} and~\eqref{eq:psimtransformation} then imply that we can rewrite the transformation of the shear under a supertranslation in a nonradiative regime of $\scri^{+}$ as
\begin{align}
\text{Re}\left[\bar{\eth}^{2}\sigma\right]'&=m'+\Psi_{\text{M}}'\nonumber\\
&=m+\left(\Psi_{\text{M}}-\eth^{2}\bar{\eth}^{2}\alpha\right).
\end{align}
This shows that by fixing the supertranslation freedom with the Moreschi supermomentum one also immediately has an understanding of how the shear will transform. Consider applying the supertranslation that maps the $\ell\geq1$, i.e., the nontemporal,
components of $\Psi_{\text{M}}$ to zero in a nonradiative regime of $\scri^{+}$.
For bound systems, which asymptote to regimes of $\scri^{+}$ that are nonradiative and stationary,\footnote{%
For bound systems, there exist frames at $u\to
\pm\infty$ where the $\ell\geq2$ components of the
Bondi mass aspect $m$ are exactly zero. While these frames at
$u\rightarrow\pm\infty$ exist, they need not be the same.}
such a transformation also maps the shear to zero in the nonradiative regime of $\scri^{+}$ because in this regime $\text{Re}\left[\bar{\eth}^{2}\sigma\right]=\Psi_{\text{M}}$.
Consequently, for systems like
BBHs, by performing such a mapping at $u\rightarrow-\infty$ or at
$u\rightarrow+\infty$, one can wholly fix the supertranslation freedom
of the system so that it agrees with PN waveforms or
QNM models.
Moreschi calls a frame in which the $\ell\geq1$ components of
$\Psi_{\text{M}}$
are zero a \emph{nice section} of $\scri^{+}$; we, like~\cite{Mitman:2021xkq,MaganaZertuche:2021syq}, will instead
call it the \emph{superrest} frame. Furthermore, since $\Psi_{\text{M}}$ in the superrest frame at $u\rightarrow\pm\infty$ is equal to Bondi mass $M_{\text{B}}$, the supertranslation $\alpha$ that maps to the superrest frame at a single time can be computed with relative ease via the implicit equation
\begin{align}
\label{eq:nicesection}
\eth^{2}\bar{\eth}^{2}\alpha=\Psi_{\text{M}}(u=\alpha,\theta,\phi)+k_{\text{rest}}(\alpha,\theta,\phi)^{3}M_{\text{B}}(\alpha).
\end{align}
This equation for $\alpha$ is obtained by taking
$\Psi_{\text{M}}'=M_{\text{B}}$ in Eq.~\eqref{eq:psimtransformation}
and rearranging terms. Note that here the
conformal factor on the right, $k_{\text{rest}}$, is a special case of
the conformal factor given in Eq.~\eqref{eq:conformalfactor}: namely
the one coming from a boost to the instantaneous rest
frame at some fixed time, computed via
\begin{align}
k_{\text{rest}}(u,\theta,\phi)\equiv\frac{1}{\gamma\left(1-\vec{v}\cdot\vec{r}\right)}=\frac{M_{B}}{P_{a}n^{a}}.
\end{align}
Apart from this, Dain and Moreschi also proved that, provided a certain condition on the energy flux, which is always obeyed by nonradiative regimes of $\scri^{+}$, this equation always has a regular solution~\cite{Dain:2000lij}. Therefore, for the frames that we are mapping to, such a supertranslation will always exist. Furthermore, Dain and Moreschi also showed that Eq.~\eqref{eq:nicesection} can be solved iteratively. That is, if one wishes to find the supertranslation that maps the system to the superrest frame at time $u_{0}$, they can take Eq.~\eqref{eq:nicesection} and evaluate the right-hand side at time $u=u_{0}$, solve for $\alpha$, evaluate the right-hand side at time $u=\alpha$, solve for a new $\alpha$, etc., until $\alpha$ converges to a solution. We will make use of this fact in Sec.~\ref{sec:numerics}.

What remains unclear at this point, however, is whether the
Moreschi supermomentum is also the right charge to use for
fixing the frame of systems with unbound masses. For these
unbound systems, the $\ell\geq2$ components of $m$ can have a nonzero net change between $u\rightarrow\pm\infty$,
which means that mapping the $\ell\geq1$ components of $\Psi_{\text{M}}$ to zero need not map the system
to a shear-free section of $\scri^{+}$. Despite this, it may be
that shear-free sections are not necessarily that
meaningful and really what one should strive for when fixing the
supertranslation freedom is ensuring that the shear has only hard
contributions~\cite{Strominger:2013jfa}, i.e., no contribution from the energy flux. In this work, however, because we focus on bound systems, like BBHs and perturbed BHs, it will suffice to only consider the Moreschi
supermomentum as the charge for fixing the supertranslation freedom of our systems.

\section{PN supermomentum}
\label{sec:pnsupermomentum}

With the Moreschi supermomentum now identified as the
supertranslation charge that can map BBH systems
(or other bound systems) to shear-free sections of $\scri^{+}$, there are two obvious BMS frames
that one can map to: the superrest frame at either
$u\rightarrow-\infty$ or $u\rightarrow+\infty$. The first choice can be
understood as mapping the system to the same frame as PN waveforms,
i.e., shear-free in the infinite past. The second is naturally
understood as mapping to the superrest frame of the remnant BH, i.e.,
making the metric equivalent to the Kerr metric rather than a
supertranslated Kerr metric~\cite{Compere:2016hzt}.
Waveforms produced by numerical relativity, however, are finite in time
and thus do not contain information about $u\rightarrow-\infty$
or $u\rightarrow+\infty$. Thus, for fixing the frame of
these waveforms we need to know what we should be mapping the
numerical Moreschi supermomentum to with the Bondi times that we
have access to. For mapping to the superrest frame of the remnant BH
($u\rightarrow+\infty$), this is
simple because the
radiation decays fast enough that mapping at or near the end of the simulation is a reasonable approximation to $u\rightarrow+\infty$. For mapping to the same
frame as PN waveforms, however, we cannot rely on approximations
because the radiation during the inspiral of a BBH merger cannot be considered
negligible. We instead need to know the Moreschi supermomentum predicted by PN theory. This will allow us to map the numerical supermomentum to equal that of the PN system, which serves as a proxy for mapping our NR system to the superrest frame at spacelike infinity.
Accordingly, we now perform a PN calculation of the Moreschi supermomentum.\\

\subsection{PN Moreschi supermomentum}

The main ingredients for this PN calculation are the
PN expressions for the strain both with
and without the BH spins~\cite{Blanchet:2013haa,Favata:2008yd,Boyle:2014ioa,Buonanno:2012rv},\footnote{Note that there was a mistake made in~\cite{Boyle:2014ioa} when calculating the spin-spin terms at 2PN order. We have corrected this mistake prior to using the PN strain in our calculations that follow.} the orbital energy~\cite{Blanchet:2013haa,Bohe:2015ana,Marsat:2014xea}, and the luminosity~\cite{Blanchet:2013haa,Bohe:2015ana,Marsat:2014xea}.
Using Eq.~\eqref{eq:conversion} to replace the shear
by the strain in Eq.~\eqref{eq:psimenergyflux} and then moving the ADM mass to the other side of the equation yields
\begin{align}
\label{eq:psimpn}
\Psi_{\text{M}}+M_{\text{ADM}}=\int_{-\infty}^{u}|\dot{\sigma}|^{2}du=\frac{1}{4}\int_{-\infty}^{u}|\dot{h}|^{2}du,
\end{align}
Therefore, provided
a spherical harmonic decomposition of the
PN strain, any spherical harmonic mode of the supermomentum can
be computed by integrating the product of various spin-weighted
spherical harmonics over the two-sphere as well as integrating the
products of various modes of the strain with respect to time:
\begin{align}
\Psi_{\text{M}}^{(\ell,m)}&=\frac{1}{4}\sum\limits_{\ell_{1},|m_{1}|\leq\ell_{1}}\sum\limits_{\ell_{2},|m_{2}|\leq\ell_{2}}\nonumber\\
&\phantom{=.}\left[\int_{S^{2}}\overline{\phantom{}_{0}Y_{\ell,m}}\,\overline{\phantom{}_{-2}Y_{\ell_{1},m_{1}}}\,\phantom{}_{-2}Y_{\ell_{2},m_{2}}d\Omega\right]\nonumber\\
&\phantom{=.}\left[\int\overline{\dot{h}^{(\ell_{1},m_{1})}}\,\dot{h}^{(\ell_{2},m_{2})}du\right].
\label{eq:Moreschimode}
\end{align}
The first of these integrals can be easily computed from the
spin-weighted spherical harmonics' relationship to the Wigner
$D$-matrices combined with the known integral of a product of three $D$ matrices~\cite{Campbell:1970ww}, which produces
\begin{widetext}
\begin{align}
\int_{S^{2}}\phantom{}_{s_{1}}Y_{\ell_{1}m_{1}}\,\phantom{}_{s_{2}}Y_{\ell_{2}m_{2}}\,\phantom{}_{s_{3}}Y_{\ell_{3}m_{3}}d\Omega=\sqrt{\frac{(2\ell_{1}+1)(2\ell_{2}+1)(2\ell_{3}+1)}{4\pi}}\begin{pmatrix}\ell_{1}&\ell_{2}&\ell_{3}\\m_{1}&m_{2}&m_{3}\end{pmatrix}\begin{pmatrix}\ell_{1}&\ell_{2}&\ell_{3}\\-s_{1}&-s_{2}&-s_{3}\end{pmatrix}.
\end{align}
\end{widetext}
For this part of the calculation, we also need the identity
\begin{align}
\overline{\phantom{}_{s}Y_{\ell,m}}=(-1)^{s+m}\phantom{}_{-s}Y_{\ell,-m}.
\end{align}

Next consider the integral with respect to time
in the final part of Eq.~\eqref{eq:Moreschimode}.
Because any mode of the strain can be written as
\begin{align}
h^{(\ell,m)}=\frac{2M\nu x}{R}\sqrt{\frac{16\pi}{5}}\mathcal{H}^{(\ell, m)}e^{-im\psi},
\end{align}
where $M\equiv m_{1}+m_{2}$, $\nu\equiv m_{1}m_{2}/M^{2}$,
$x\equiv(M\omega)^{2/3}$ is the usual PN parameter, $\psi$ is the
auxiliary phase variable (see Eq.~(321) of
\cite{Blanchet:2013haa}),
and $\mathcal{H}^{(\ell, m)}$ is a polynomial in $x$, we can perform a change of coordinates from $u$ to $x$ to obtain a series of integrals of the form
\begin{align}
\int_{x_{0}}^{x}\dot{x}\left(d\psi/dx\right)^{a}x^{b}e^{-iA\psi(x)}dx,
\end{align}
where $a\in\{0,1,2\}$, $b\in\mathbb{N}$, $A\in\mathbb{Z}$, and $x_{0}$ corresponds to an arbitrary initial frequency $\omega_{0}$.
To evaluate these various integrals over $x$, consider first integrating the following expression by parts to obtain
\begin{align}
&\int_{x_{0}}^{x}f(x)e^{\varphi(x)}dx=\nonumber\\
&\left[\left(\frac{d\varphi}{dx}\right)^{-1}fe^{\varphi}\right]_{x_{0}}^{x}\nonumber\\
&-\int_{x_{0}}^{x}\left[\left(\frac{d\varphi}{dx}\right)^{-1}\frac{df}{dx}-\left(\frac{d\varphi}{dx}\right)^{-2}\frac{d^{2}\varphi}{dx^2}f\right]e^{\varphi}dx.
\end{align}
Because $\varphi\propto\psi\sim x^{-5/2}$, the PN order of the integrand on
the right-hand side ends up being 2.5PN higher than the original integrand.
Thus we can evaluate these integrals by integrating
by parts until the unevaluated integrals have been pushed
to a PN order above what we consider.
Alternatively, one can evaluate such integrals with either the stationary
phase approximation or the method of steepest
descent~\cite{bender1999advanced}, but the result should be the
same. By carrying
out this integration procedure, we then find that we can compute the PN
Moreschi supermomentum to relative 3PN order when spins are not included
and relative 2PN order when spins are included, with the limiting factor
being the available PN order of the strain that we input into Eq.~\eqref{eq:psimpn}. We write the modes of the
Moreschi supermomentum as
\begin{align}
  \label{eq:Psi_M_decomp_calP}
\Psi_{\text{M}}^{(\ell,m)}=\frac{2M\nu x}{R^{2}}\sqrt{\frac{\pi}{4}}\mathcal{P}^{(\ell,m)}e^{-im\psi}-M\delta_{\ell,0}\delta_{m,0}
\end{align}
where $\mathcal{P}^{(\ell,m)}$ is a polynomial in $x$.
The term in $M$ that
arises when $(\ell,m)=(0,0)$ is because of the presence of the ADM
mass in Eq.~\eqref{eq:psimenergyflux}. While we write our results in
full in Appendices~\ref{sec:appendixa} and~\ref{sec:appendixb}, we
provide a few of the most interesting modes below. These results have
been obtained using \texttt{Mathematica}.\footnote{The authors are willing to share the \texttt{Mathematica} notebook used for this calculation upon reasonable request.}

\begin{widetext}
	\begin{subequations}
		\allowdisplaybreaks{}
		\begin{align}
			\label{eq:PNMoreschisupermomentum_subset}
			\mathcal{P}^{(0,0)}&=1+x \left(-\frac{3}{4}-\frac{\nu }{12}\right)+x^2 \left(-\frac{27}{8}+\frac{19 \nu }{8}-\frac{\nu ^2}{24}\right)+x^3
			\left(-\frac{675}{64}+\left(\frac{34445}{576}-\frac{205 \pi ^2}{96}\right) \nu -\frac{155 \nu ^2}{96}-\frac{35 \nu ^3}{5184}\right),\\
			\mathcal{P}_{\text{spin}}^{(0,0)}&=x^{3/2} \left(\frac{14 S_\ell}{3 M^2}+\frac{2 \delta  \Sigma _\ell}{M^2}\right)\nonumber\\
			&\phantom{=.}+x^2 \left(-\frac{16 \vec{S}\cdot\vec{S}+3 \vec{\Sigma}\cdot\vec{\Sigma} +32 S_\ell^2+9 \Sigma _\ell^2}{12 M^4}-\frac{4 \delta  \left(\vec{S}\cdot\vec{\Sigma}+2 S_\ell \Sigma _\ell\right)}{3 M^4}+
			\frac{4 \left(\vec{\Sigma}\cdot\vec{\Sigma} +2 \Sigma _\ell^2\right)\nu}{3 M^4}\right),\\
			\mathcal{P}^{(1,1)}&=\frac{43}{70}\sqrt{\frac{3}{2}}\Bigg\lbrace x^3\left(\frac{1856\delta\nu}{129}\right)\Bigg\rbrace,\\
			\mathcal{P}_{\text{spin}}^{(1,1)}&=0,\\
			\mathcal{P}^{(2,0)}&=\frac{2}{7}\sqrt{5}\Bigg\lbrace1+x \left(-\frac{4075}{4032}+\frac{67 \nu }{48}\right)+x^2
			\left(-\frac{151877213}{67060224}-\frac{123815 \nu }{44352}+\frac{205 \nu ^2}{352}\right)+\pi x^{5/2} \left(-\frac{253 }{336}+\frac{253  \nu }{84}\right)\nonumber\\
			&\phantom{=.\frac{2}{7}\sqrt{5}\Bigg\lbrace}+x^3
			\left(-\frac{4397711103307}{532580106240}+\left(\frac{700464542023}{13948526592}-\frac{205 \pi ^2}{96}\right) \nu +\frac{69527951 \nu ^2}{166053888}+\frac{1321981
				\nu ^3}{5930496}\right)\Bigg\rbrace,\\
			\mathcal{P}_{\text{spin}}^{(2,0)}&=\frac{2}{7}\sqrt{5}\Bigg\lbrace x^{3/2} \left(\frac{16 S_\ell}{3 M^2}+\frac{419 \delta  \Sigma _\ell}{160 M^2}\right)\nonumber\\
			&\phantom{=.\frac{2}{7}\sqrt{5}\Bigg\lbrace}+x^2
			\left(-\frac{128 \vec{S}\cdot\vec{S}+24 \vec{\Sigma}\cdot\vec{\Sigma} +256 S_\ell^2+75
				\Sigma _\ell^2}{96 M^4}-\frac{4 \delta  \left(\vec{S}\cdot\vec{\Sigma} +2 S_\ell \Sigma _\ell\right)}{3 M^4}+\frac{4 \left(\vec{\Sigma}\cdot\vec{\Sigma} +2 \Sigma _\ell^2\right)\nu}{3 M^4}\right)\Bigg\rbrace,\\
			\mathcal{P}^{(3,1)}&=\frac{223}{120\sqrt{21}}\Bigg\lbrace x^3\left(\frac{3872 \delta  \nu}{223}\right)\Bigg\rbrace,\\
			\mathcal{P}_{\text{spin}}^{(3,1)}&=0.
		\end{align}
	\end{subequations}
\end{widetext}
\begin{table*}
	\label{tab:runs}
	\centering
	\renewcommand{\arraystretch}{1.2}
	\begin{tabular}{@{}l@{\hspace*{7mm}}c@{\hspace*{7mm}}c@{\hspace*{7mm}}c@{}c@{}c@{}c@{\hspace*{7mm}}c@{}c@{}c@{}c@{}}
		\Xhline{3\arrayrulewidth}
		Name & CCE radius & $q$ & $\chi_{A}$:\, & $(\hat{x},\,$ & $\hat{y},\,$ & $\hat{z})$ & $\chi_{B}$:\, & $(\hat{x},\,$ & $\hat{y},\,$ & $\hat{z})$\\
		\hline
		\texttt{q1\_nospin} & 292 & $1.0$ & & $(0,\,$ & $0,\,$ &  $0)$ & & $(0,\,$ & $0,\,$ & $0)$ \\
		\texttt{q1\_aligned\_chi0\_2} & 261 & $1.0$ & & $(0,\,$ & $0,\,$ & $0.2)$ & & $(0,\,$ & $0,\,$ & $0.2)$ \\
		\texttt{q1\_aligned\_chi0\_4} & 250 & $1.0$ & & $(0,\,$ & $0,\,$ & $0.4)$ & & $(0,\,$ & $0,\,$ & $0.4)$ \\
		\texttt{q1\_aligned\_chi0\_6} & 236 & $1.0$ & & $(0,\,$ & $0,\,$ & $0.6)$ & & $(0,\,$ & $0,\,$ & $0.6)$ \\
		\texttt{q1\_antialigned\_chi0\_2} & 274 & $1.0$ & & $(0,\,$ & $0,\,$ & $0.2)$ & & $(0,\,$ & $0,\,$ & $-0.2)$ \\
		\texttt{q1\_antialigned\_chi0\_4} & 273 & $1.0$ & & $(0,\,$ & $0,\,$ & $0.4)$ & & $(0,\,$ & $0,\,$ & $-0.4)$ \\
		\texttt{q1\_antialigned\_chi0\_6} & 270 & $1.0$ & & $(0,\,$ & $0,\,$ & $0.6)$ & & $(0,\,$ & $0,\,$ & $-0.6)$ \\
		\texttt{q1\_precessing} & 305 & $1.0$ & & $(0.487,\,$ & $0.125,\,$ & $-0.327)$ & & $(-0.190,\,$ & $0.051,\,$ & $-0.227)$ \\
		\texttt{q1\_superkick} & 270 & $1.0$ & & $(0.6,\,$ & $0,\,$ & $0)$ & & $(-0.6,\,$ & $0,\,$ & $0)$ \\
		\texttt{q4\_nospin} & 235 & $4.0$ & & $(0,\,$ & $0,\,$ & $0)$ & & $(0,\,$ & $0,\,$ & $0)$ \\
		\texttt{q4\_aligned\_chi0\_4} & 222 & $4.0$ & & $(0,\,$ & $0,\,$ & $0.4)$ & & $(0,\,$ & $0,\,$ & $0.4)$ \\
		\texttt{q4\_antialigned\_chi0\_4} & 223 & $4.0$ & & $(0,\,$ & $0,\,$ & $0.4)$ & & $(0,\,$ & $0,\,$ & $-0.4)$ \\
		\texttt{q4\_precessing} & 237 & $4.0$ & & $(0.487,\,$ & $0.125,\,$ & $-0.327)$ & & $(-0.190,\,$ & $0.051,\,$ & $-0.227)$ \\
		\texttt{SXS:BBH:0305} (GW150914) & 267 & $1.221$ && $(0,\,$ & $0,\,$ & $0.330)$ & & $(0,\,$ & $0,\,$ & $-0.440)$ \\
		\Xhline{3\arrayrulewidth}
	\end{tabular}
	\caption{Parameters of the BBH mergers used in our results. The mass ratio is $q=M_A/M_B$ and the initial dimensionless spins of the two black holes are $\chi_A$ and $\chi_B$. These simulations have been made publicly available at~\cite{ExtCCECatalog,SXSCatalog}.}
\end{table*}
For the terms in these expressions that include spins, with $M_{1}$ and $M_{2}$ and $\vec{S}_{1}$ and $\vec{S}_{2}$ the masses and spins of the two black holes,
\begin{align}
\vec{S}\equiv\vec{S}_{1}+\vec{S}_{2}
\end{align}
is the total spin vector,
\begin{align}
\vec{\Sigma}\equiv M\left(\frac{\vec{S}_{2}}{M_{2}}-\frac{\vec{S}_{1}}{M_{1}}\right)
\end{align}
can be viewed as an effective antisymmetric spin vector,
$\delta\equiv\left(M_{1}-M_{2}\right)/M$, $\hat{n}$ is the unit vector pointing from black hole $2$ to black hole
$1$, $\hat{\lambda}$ is the unit vector in the direction of
$d\hat{n}/du$, and $\hat{\ell}=\hat{n}\times\hat{\lambda}$. We
include these modes here for the following reasons. The $(0,0)$
mode is proportional to the energy radiated to future null infinity,
and matches the orbital energy results of~\cite{Blanchet:2013haa,Bohe:2015ana,Marsat:2014xea,Isoyama:2017tbp}
to 3PN order without spins and also to 2PN order with spins.
The $(1,1)$ mode corresponds to the radiated momentum
and therefore highlights that at 3PN order without spins the center-of-mass
is not stationary but oscillates about the origin. When
spins are not included, the expression for
the $(2,0)$ mode, which is the main memory mode, recovers the previous PN memory result
of~\cite{Favata:2008yd}.\footnote{Actually our result is only
proportional to the earlier result of~\cite{Favata:2008yd}. The two
differ by a factor of $\frac{1}{2}\sqrt{6}$, which comes
from the factor of $\frac{1}{2}$ needed to change the
shear to the strain and the factor of $\sqrt{6}$ that arises when
applying the $\eth^{2}$ operator to an $\ell=2$ spin-weight $-2$
spherical harmonic mode.} But, when spins are included, we observe
new memory terms that are proportional to spin.
This is a new result in
PN theory, even though this behavior has been known in the
numerical community~\cite{Pollney:2010hs,Mitman:2020pbt}. Last,
the $(3,1)$ mode highlights a new identification in post-Newtonian theory,
namely the existence of oscillatory memory modes, which arise at
3PN order and have been known to exist but have not been computed in a memory context, even though the 3PN strain without spins is complete~\cite{Favata:2008yd,Blanchet:2004ek}.

\section{Numerical analysis}
\label{sec:numerics}

With the charges needed for fixing the BMS frame summarized
in Sec.~\ref{sec:bmstransformations}, we now present numerical results of
mapping BBH waveforms to either the PN
BMS frame or the superrest frame of the remnant black hole at $i^{+}$. We numerically
evolved a set of 14 binary
black hole mergers with varying mass ratios and spin configurations using
the spectral Einstein code (SpEC)~\cite{SpECCode}. We list the
important parameters of these various BBH systems in
Table~\ref{tab:runs}. Each simulation contains roughly 19 orbits prior
to merger and is evolved until the waves from ringdown leave the
computational domain. Unlike the evolutions in the SXS catalog~\cite{SXSCatalog},
the full set of Weyl scalars has been extracted from these
runs. The waveforms have been computed using both the extrapolation
technique described in~\cite{Iozzo:2020jcu} and the
Cauchy-characteristic extraction (CCE) procedure outlined
in~\cite{Moxon:2020gha,Moxon:2021gbv}. Extrapolation is performed with the
python module \texttt{scri}~\cite{scri_url, Boyle:2013nka, Boyle:2015nqa, Boyle:2014ioa} and CCE is run with SpECTRE's CCE
module~\cite{Moxon:2020gha,Moxon:2021gbv,CodeSpECTRE}.

For the CCE extractions, the four world tubes that are available have
radii that are equally spaced between $2\lambdabar_{0}$ and
$21\lambdabar_{0}$, where $\lambdabar_0\equiv 1/\omega_0$ is the initial
reduced gravitational wavelength as determined by the orbital
frequency of the binary from the initial data. Based on the recent
work of~\cite{Mitman:2020bjf}, however, we choose to use only the
waveforms that correspond to the world tube with the second-smallest
radius, since these waveforms have been shown to minimally violate the
Bianchi identities. For clarity, we provide the world tube radius used
for each system in Table~\ref{tab:runs}. All of these 14 BBH systems'
waveforms have been made publicly available
at~\cite{ExtCCECatalog,SXSCatalog}.

\figPNBMScomparison

As mentioned above, the asymptotic strain waveforms are created using
two methods: extrapolation and CCE. The first method utilizes
Regge-Wheeler-Zerilli (RWZ) extraction to compute the strain waveform on a
series of concentric spheres of constant coordinate radius and then
extrapolates
these values to
future null infinity $\scri^{+}$ by fitting a power series in $1/r$~\cite{Sarbach:2001qq, Regge:1957td, Zerilli:1970se, Boyle:2019kee, Iozzo:2020jcu, Boyle:2009vi}. This is the strain found in the SXS
catalog. The CCE method,
which is more faithful, instead
uses world tube data from a
Cauchy evolution as the inner boundary data for a nonlinear evolution
of the Einstein field equations on null
hypersurfaces extending to
$\scri^{+}$~\cite{Moxon:2020gha,Moxon:2021gbv}. CCE requires freely
specifying the strain on the initial null hypersurface of the simulation. As in~\cite{Mitman:2020pbt, Mitman:2020bjf, Mitman:2021xkq,MaganaZertuche:2021syq}, we
choose this field to match the value and the first radial derivative
of $h$ from the Cauchy data on the world tube using the ansatz
\begin{align}
h(u=0,r,\theta^{A})=\frac{A(\theta^{A})}{r}+\frac{B(\theta^{A})}{r^{3}},
\end{align}
where the two coefficients $A(\theta^{A})$ and $B(\theta^{A})$ are
fixed by the Cauchy data on the world tube.

Lastly, when we compute BMS charges and transform our asymptotic variables to either the PN BMS frame or the superrest frame, we use the code
\texttt{scri}~\cite{scri_url, Boyle:2013nka, Boyle:2014ioa, Boyle:2015nqa}, specifically the function \texttt{map\_to\_superrest\_frame}.

\subsection{Fixing the BMS frame} 
\label{sec:bmsframefixing}

\figsuperrestcomparison

For fixing the Poincar\'e freedom of our systems,
Sec.~\ref{sec:bmstransformationsPoincare} pointed out that the
ideal charge for fixing the translation and boost transformations
is the center-of-mass charge, Eq.~\eqref{eq:CoMcharge}, and
the ideal charge for fixing the rotation would
typically be the usual angular momentum charge. However, because we want
to map numerical waveforms to the PN BMS frame, for which
the $\Psi_{1}$ and $\Psi_{2}$ Weyl scalars are unknown, for this mapping it is more convenient to use
the charge that corresponds to the rotations on future null
infinity, Eq.~\eqref{eq:angularvelocityvector}. Meanwhile, for mapping to superrest frame we use the spin vector, Eq.~\eqref{eq:spincharge}, because when we are in the center-of-mass frame of the remnant black hole there is no orbital angular momentum, so we need only care about the spin contribution. Apart from these Poincaré freedoms,
Sec.~\ref{sec:bmstransformationsupertranslations} illustrated that for the supertranslation freedom
the Moreschi supermomentum can be used to map NR waveforms for comparison to either PN waveforms or QNM models. Thus, the entire BMS freedom of 
the system can be fixed via these charges: the center-of-mass charge,
the rotation charge, and the Moreschi supermomentum.

For our NR systems, however, we find that obtaining BMS transformations from these charges must be done iteratively
to ensure the convergence of the process.\footnote{It may be that this need to find these transformations iteratively is true in the analytical case as well, but this is beyond the scope of this paper.} Consequently,
we fix the BMS frame as follows:
\begin{enumerate}[I.]
	\item Find the space translation and boost that minimize the
	center-of-mass charge $G^{a}$ over a large window; i.e., compute
	the center-of-mass charge and fit it with a linear function in time. The boost is
the slope and the space translation is the intercept.
	\item Find the proper supertranslation that maps the
	$\ell\geq2$ components of $\Psi_{\text{M}}$
	to the values obtained by PN (PN BMS frame) or to $M_{B}$
	(superrest frame); i.e., compute the supertranslation using
	Eq.~\eqref{eq:psimtransformation} where $\Psi_{\text{M}}'$
	is either $\Psi_{\text{M}}^{\text{PN}}$ or $M_{\text{B}}$. More practically, for fixing the frame using data at time $u=u_{0}$, we solve Eq.~\eqref{eq:psimtransformation} for $\eth^{2}\bar{\eth}^{2}\alpha$ by taking $\alpha=u_{0}$ on the right-hand side, compute an approximate $\alpha$ by inverting $\eth^{2}\bar{\eth}^{2}$, and then iterate this procedure by taking $\alpha$ on the right-hand side to be the $\alpha$ obtained from the prior iteration until $\alpha$ converges.
\item Apply the supertranslation (the space translation and the proper supertranslation) to the original asymptotic quantities.
	\item Compute the rotation that maps the rotation charge (either
	the charge in Eq.~\eqref{eq:angularvelocityvector} for the PN BMS
	frame or in Eq.~\eqref{eq:spincharge} for the superrest frame)
	to the values computed using a PN waveform (PN BMS
	frame) or to be parallel to the $+\hat{z}$-axis (superrest
	frame). We do this calculation using Davenport's solution to Wahba's problem
	to find the quaternion that best aligns the two
	charge vectors (see Sec.~5.3 of~\cite{Markley_2014}).
	\item Apply both the supertranslation and rotation to the original asymptotic quantities.
	\item Repeat step I to obtain a new space translation and boost transformation for the transformed quantities.
	\item Apply the center-of-mass transformation to the transformed asymptotic quantities.
	\item If mapping to the PN BMS frame, perform a time/phase alignment using a 2D minimization of the error between the NR and PN strain waveforms.
\end{enumerate}

When finding these BMS transformations from the BMS charges,
we find them iteratively. Put differently, we solve for the transformation,
transform the charge, solve for the transformation again, etc.,
until the transformation we are computing converges. This typically happens within
five iterations, which matches the results of~\cite{Mitman:2021xkq}
for the center-of-mass transformations.  Note that the reason
why we apply a space translation and boost after solving for the
supertranslation and rotation and transforming the asymptotic
quantities is because we find that doing so is necessary to
obtain the behavior that we expect from the center-of-mass
transformation. Empirically we found that to minimize the new center-of-mass-charge, we must apply a space translation and boost after solving for the supertranslation and rotation, and transforming the asymptotic quantities.
  
As for the values of the transformations output by this charge-based frame fixing method, we find the following. The center-of-mass transformation is consistent with the work of~\cite{Mitman:2021xkq}. The rotation quaternion tends to be close to the unit quaternion, except when working with systems that are precessing, in which case it is hard to predict. The modes of the supertranslation tend to be nearly zero when $m\not=0$ and fairly nontrivial when $m=0$. But this is simply because the
$m=0$ memory modes dominate in PN theory, so we must correspondingly
apply a supertranslation with larger $m=0$ coefficients.

All of the code used to perform these computations and frame transformations has been incorporated into the open-source code \texttt{scri}~\cite{scri_url, Boyle:2013nka, Boyle:2014ioa, Boyle:2015nqa}, which has been validated by unit tests and prior works~\cite{scri_url, Boyle:2013nka, Boyle:2014ioa, Boyle:2015nqa,Mitman:2021xkq}.

\subsection{Comparison of charge to optimizer method}
\label{sec:comparisonprevresults}

\figPNcomparison

\figQNMcomparison

Here we compare
the charge-based scheme for fixing the BMS frame to the
previous method, which relied on optimization algorithms.
First, in Fig.~\ref{fig:PNBMScomparison} we show the relative
error between the NR and PN strain waveforms once the
NR system is mapped to the PN BMS frame. For reference,
we show
the relative error between these waveforms when
no frame fixing is performed, or when only a time and phase fixing is
performed by minimizing the absolute error between the NR and
PN waveforms.\footnote{The main reason why the time and phase
fixing produces such poor results is because the waveforms being
compared contain memory effects, which require a supertranslation
to be aligned.} Both methods compute the relative error between
these two waveforms over a three-orbit window during the early
inspiral phase and over the whole two-sphere for every spin-weighted
spherical harmonic mode up to $\ell_{\text{max}}=8$. Explicitly,
\begin{align}
\text{rel. err.}\left(h^{\text{NR}},h^{\text{PN}}\right)=||h^{\text{NR}}-h^{\text{PN}}||/||h^{\text{PN}}||,
\end{align}
where
\begin{align}
\label{eq:norm}
||h||\equiv\sqrt{\int_{S^{2}}\left(\int_{u_{1}}^{u_{2}}\left|h(\ell_{\text{min}}\leq\ell\leq\ell_{\text{max}})\right|^{2}du\right)d\Omega}.
\end{align}
Here $u_{1}$ is
the time that is $1200M$ past the beginning of the
NR simulation, $u_{2}$ the time three-orbits beyond $u_{1}$,
$\ell_{\text{min}}=2$, and $\ell_{\text{max}}=8$. As can be seen
by comparing the plain green bars (charge-based method) to
the patterned green bars (optimization algorithm)
throughout Fig.~\ref{fig:PNBMScomparison},
the charge-based frame fixing method produces an error that
is nearly identical to the previous method
for every BBH system, but with a typical run time of 30 minutes instead of 12
hours: a speedup of $\sim 24\times$.

In Fig.~\ref{fig:superrestcomparison} we show the norm of the
$\ell>0$ components of $\Psi_{\text{M}}$ during the final
$100M$ of the numerical simulation, which is roughly $200M$ past the
peak of the $L^{2}$ norm of the strain. We calculate this peak time by finding the time at which the
square of Eq.~\eqref{eq:norm}, without the time integration, reaches
its maximum value. As can be seen by comparing the plain green
bars (charge-based method) to the patterned green bars (optimization
algorithm), the new method produces
errors that are roughly two orders of
magnitude better than the optimization algorithm used in~\cite{Mitman:2021xkq,MaganaZertuche:2021syq}. Thus,
it is evident that fixing the BMS frame to be the superrest frame
of the remnant black hole is remarkably improved when using the new, charge-based frame fixing method. This improvement is primarily due to our
ability to obtain more physically motivated supertranslations from
the Moreschi supermomentum via Eq.~\eqref{eq:nicesection}, rather than an optimization algorithm
that need not produce physical results.

\subsection{Comparison of NR to PN and QNMs}
\label{sec:comparison}

Thus far, we have shown
why fixing the BMS frame of NR
systems is important for comparing NR waveforms to PN waveforms
or for modeling the ringdown phase of NR waveforms with QNMs.
We now show why CCE waveforms and BMS frame
fixing will help usher in the next generation of NR waveforms. 
We do so by comparing both CCE and extrapolated waveforms to PN waveforms in
Fig.~\ref{fig:PNcomparison} and QNMs in Fig.~\ref{fig:QNMcomparison}.

In Fig.~\ref{fig:PNcomparison}, like Fig.~\ref{fig:PNBMScomparison},
we show the relative error between a NR and a PN strain waveform
over a three-orbit window. In the top panel, we compute the relative error between
these waveforms using the $(2,2)$, $(2,1)$, $(3,3)$, $(3,2)$,
$(3,1)$, $(4,4)$, $(4,3)$, $(4,2)$, and $(5,5)$ modes.
These are typically the most important modes without the $m=0$ modes.
This collection of modes also matches those used in the \texttt{NRHybSur3dq8} surrogate~\cite{Varma:2018mmi}.
In the bottom panel, we instead compute the relative error using every
mode up to $\ell=8$. The error obtained when using CCE waveforms is shown in
green, while the error when using extrapolated waveforms is shown in
red. Furthermore, for the CCE waveforms we show two errors: one where
we perform a time/phase alignment (faint) and one where we perform a
BMS frame alignment (full). For the extrapolated waveforms, however,
the error is computed only with a time/phase alignment,
since this is what has been used before Ref.~\cite{Mitman:2021xkq}.\footnote{We should also note that this
charge-based method for fixing the BMS frame cannot be performed on
the vast majority of the publicly available extrapolated waveforms
in the SXS Catalog~\cite{SXSCatalog}, since their Weyl scalars
have not been extracted.} As can be seen throughout the top panel,
where the $m=0$ modes have been excluded, provided that we have fixed the BMS
frame of the CCE waveforms, then these waveforms are on a par with
the extrapolated waveforms.\footnote{The reason why the faint bars
produce such poor errors is because the CCE waveforms are output
in an arbitrary BMS frame and thus require a supertranslation to
obtain sensible results.} However, if we include the $m=0$ modes
as in the bottom panel, then we find that the CCE waveforms are
easily able to outperform the extrapolated waveforms for every
type of system considered. This is because CCE waveforms, unlike
extrapolated waveforms, contain memory effects that are also
present in the PN treatment.

For working in the superrest frame, in Fig.~\ref{fig:QNMcomparison},
we show the mismatch between a NR waveform and a QNM model
constructed from 100 QNM modes that are chosen using the ranking
system presented in~\cite{MaganaZertuche:2021syq}.\footnote{While
there is an ongoing debate within the QNM community concerning the
possible over-fitting of QNMs to NR waveforms, because we are only
comparing fits to different waveforms rather than the fits themselves
this is not important to our results.} In the top panel we show the
mismatch between the strain waveforms, while in the bottom panel we
show the mismatch between the news waveforms, where the news waveform
is just the time derivative of the strain waveform. The mismatch when
using CCE waveforms is shown in green, while the mismatch when using
extrapolated waveforms is shown in red. We compute the mismatch via
\begin{align}
\mathcal{M}\left(h^{A},h^{B}\right)\equiv1-\Re\left[\frac{\langle h^{A},h^{B}\rangle}{\sqrt{\langle h^{A},h^{A}\rangle\langle h^{B},h^{B}\rangle}}\right],
\end{align}
where
\begin{align}
\langle h^{A},h^{B}\rangle\equiv\int_{S^{2}}\int_{u_{0}}^{\infty}h^{A}\overline{h^{B}}\,du\,d\Omega.
\end{align}
Again, for CCE we show two mismatches: one where the CCE system
is in the center-of-mass frame of the remnant black hole (faint)
and one in the superrest frame of the remnant black hole (full). As
can be seen in the top panel, mapping to the superrest frame is
essential for modeling CCE waveforms with QNMs. This is because,
unlike extrapolated waveforms, CCE waveforms contain memory effects
so the CCE waveform will not decay to zero at timelike
infinity, while the QNMs will. As a result, to model these
NR waveforms with QNMs, fixing the supertranslation freedom is vitally
important so that the waveforms do decay to zero.
However, this is not the only impact that fixing
the BMS frame has on NR waveforms. In the bottom panel, which shows
the mismatch between the news waveforms, there continues to be an
improvement by mapping to the superrest frame, even though there is
no memory in these news waveforms. This is because when applying a
supertranslation, there is also important mode mixing
from expressing the first term of~\eqref{eq:sheartransformation}
in terms of the untransformed time, i.e.,
\begin{align}
\sigma(u')=\sum\limits_{n=0}^{\infty}\frac{1}{n!}\left[\left(k(u-\alpha)-u\right)\frac{\partial}{\partial u}\right]^{n}\sigma(u).
\end{align}
For more on this, see
Fig. 7 of~\cite{MaganaZertuche:2021syq} and the related text. Lastly,
we should also note that by comparing Fig.~\ref{fig:QNMcomparison}
to Fig. 10 of~\cite{MaganaZertuche:2021syq}, one can see that the
mismatches obtained are practically identical, which means that the
results of~\cite{MaganaZertuche:2021syq} should not be
impacted by this new scheme for mapping to the superrest frame
using BMS charges.

\section{Discussion}

We have presented a new procedure for fixing the entire
BMS frame of the data produced by
NR simulations.
The method relies on the use of
BMS charges rather than optimization algorithms, like those used in~\cite{Mitman:2021xkq,MaganaZertuche:2021syq}. This
charge-based frame fixing method fixes the system's frame using the center-of-mass
charge (Eq.~\eqref{eq:CoMcharge}) and the rotation charge
(Eq.~\eqref{eq:angularvelocityvector}) for the Poincar\'e freedoms, and the Moreschi
supermomentum $\Psi_{\text{M}}$ (Eq.~\eqref{eq:Moreschisupermomentum}) for the supertranslation freedom. If the
time and phase freedoms need to be fixed, e.g., for comparing NR systems
to PN, then a 2D minimization of the error between NR and PN strain waveforms is performed. This code
has been made publicly available in the python module
\texttt{scri}~\cite{scri_url,Boyle:2013nka,Boyle:2014ioa,Boyle:2015nqa}.

The BMS transformations are obtained by finding the transformation that
changes the corresponding charge in a prescribed way. For example, to map to the system's center-of-mass frame we find the transformation that maps the center-of-mass
charge to have an average of zero. Accordingly,
the transformations can be found much faster than
if they were computed with a minimization scheme. For the BBH systems
and frames we considered, we found that this new charge-based
method converges in roughly $30$ minutes,
rather than the $12$ or more hours that are needed by the previous optimization algorithm.

In particular, using the charge-based frame fixing, we mapped
14 binary systems to the PN BMS frame, i.e., the frame that PN
waveforms are in. Apart from this,
we mapped these systems to the superrest frame, i.e., the
frame in which the metric of the remnant black hole matches the Kerr metric at timelike infinity. For mapping to the PN BMS frame, we fixed the Poincar\'e frame by
mapping the center-of-mass charge to an average of zero and
the rotation charge to match the rotation charge of the PN waveform. For the
supertranslation freedom, however, we found that it was necessary to
calculate the PN Moreschi supermomentum so that we could find
the supertranslation via Eq.~\eqref{eq:psimtransformation}, which
maps the numerical Moreschi supermomentum to the PN supermomentum
during the early inspiral phase.

In Sec.~\ref{sec:pnsupermomentum} we performed this calculation
by using the relation that writes $\Psi_{\text{M}}$
in terms of the energy flux (see, e.g.,
Eq.~\eqref{eq:psimenergyflux}).
This calculation provides $\Psi_{\text{M}}$ to 3PN
order without spins and 2PN order with spins.
Moreover, this calculation also leads to
oscillatory and spin-dependent memory terms that either have not been identified or have been missing
from the existing
PN strain expressions. These strain terms, as well as the complete
expression for the PN Moreschi supermomentum, can be found in
Appendices~\ref{sec:appendixa} and~\ref{sec:appendixb}.

In Sec.~\ref{sec:bmsframefixing} we then described our procedure for fixing the BMS frame. In
Sec.~\ref{sec:comparisonprevresults} we used our new code
to map to the PN BMS frame (see Fig.~\ref{fig:PNBMScomparison})
or to minimize the Moreschi supermomentum during ringdown, that is, map
to the superrest frame (see Fig.~\ref{fig:superrestcomparison}).
In each case, we compared
with the previous code that relied on minimizers. Overall, we found
that these two procedures tend to yield the same errors when
mapping to the PN BMS frame, while the new charge-based procedure
yields better errors when mapping to the superrest frame.

Finally, in Sec.~\ref{sec:comparison} we considered CCE waveforms
whose BMS frame has been fixed with this new method
to be either the PN BMS frame or the superrest frame of the remnant
black hole. We showed that such waveforms
can notably outperform extrapolated waveforms, which are the current waveforms in the SXS catalog. This
suggests that CCE waveforms and BMS frame fixing
will be vital in the future for performing more correct numerical relativity simulations and conducting better waveform modeling.

\label{sec:discussion}

\section*{Acknowledgments}
L.C.S. thanks Laura Bernard for insightful discussions, and the
Benasque science center and organizers of the conference ``New
frontiers in strong gravity'' for enabling these conversations.
We thank Laura Bernard, Luc Blanchet, Guillaume Faye, and Tanguy Marchand for sharing a \texttt{Mathematica} notebook that included the PN expressions from Appendix B of~\cite{Bernard:2017ktp}.
Computations for this work were performed with the Wheeler cluster at Caltech. This work was supported in part by the Sherman Fairchild Foundation and by NSF Grants No.~PHY-2011961, No.~PHY-2011968, and No.~OAC-1931266 at Caltech, as well as NSF Grants No.~PHY-1912081 and No.~OAC-1931280 at Cornell.
The work of L.C.S. was partially supported by NSF CAREER Award
PHY-2047382.

\def\bibsection{\section*{References}}

\appendix

\section{PN MORESCHI SUPERMOMENTUM}
\label{sec:appendixa}

The complete results from our PN calculation of the modes of the Moreschi supermomentum are as follows:

\begin{widetext}
\begin{subequations}
\label{eq:PNMoreschisupermomentum}
\allowdisplaybreaks{}
\begin{align}
\mathcal{P}^{(0,0)}&=1+x \left(-\frac{3}{4}-\frac{\nu }{12}\right)+x^2 \left(-\frac{27}{8}+\frac{19 \nu }{8}-\frac{\nu ^2}{24}\right)+x^3
\left(-\frac{675}{64}+\left(\frac{34445}{576}-\frac{205 \pi ^2}{96}\right) \nu -\frac{155 \nu ^2}{96}-\frac{35 \nu ^3}{5184}\right),\\
\mathcal{P}_{\text{spin}}^{(0,0)}&=x^{3/2} \left(\frac{14 S_\ell}{3 M^2}+\frac{2 \delta  \Sigma _\ell}{M^2}\right)\nonumber\\
&\phantom{=.}+x^2 \left(-\frac{16 \vec{S}\cdot\vec{S}+3 \vec{\Sigma}\cdot\vec{\Sigma} +32 S_\ell^2+9 \Sigma _\ell^2}{12 M^4}-\frac{4 \delta  \left(\vec{S}\cdot\vec{\Sigma}+2 S_\ell \Sigma _\ell\right)}{3 M^4}+
\frac{4 \left(\vec{\Sigma}\cdot\vec{\Sigma} +2 \Sigma _\ell^2\right)\nu}{3 M^4}\right),\\
\mathcal{P}^{(1,1)}&=\frac{43}{70}\sqrt{\frac{3}{2}}\Bigg\lbrace x^3\left(\frac{1856\delta\nu}{129}\right)\Bigg\rbrace,\\
\mathcal{P}_{\text{spin}}^{(1,1)}&=0,\\
\mathcal{P}^{(2,1)}&=0,\\
\mathcal{P}_{\text{spin}}^{(2,1)}&=\frac{61}{14\sqrt{30}}\Bigg\lbrace x^{3/2} \left(-\frac{\left(S_n-i S_{\lambda }\right)}{M^2}-\frac{375 \delta  \left(\Sigma _n-i \Sigma _{\lambda }\right)}{488 M^2}\right)\nonumber\\
&\phantom{=.\frac{61}{14\sqrt{30}}\Bigg\lbrace}+x^2 \Bigg(\frac{10 \left(3 S_l \left(S_n-i S_{\lambda
	}\right)+\Sigma _l \left(\Sigma _n-i \Sigma _{\lambda }\right)\right)}{61 M^4}+\frac{15 \delta 
	\left(\left(S_n-i S_{\lambda }\right) \Sigma _l+S_l \left(\Sigma _n-i \Sigma _{\lambda }\right)\right)}{61 M^4}\nonumber\\
&\phantom{=.\frac{61}{14\sqrt{30}}\Bigg\lbrace+x^{2}\Bigg(}-\frac{30 \nu  \Sigma _l \left(\Sigma _n-i \Sigma _{\lambda }\right)}{61 M^4}\Bigg)\Bigg\rbrace,\\
\mathcal{P}^{(2,0)}&=\frac{2}{7}\sqrt{5}\Bigg\lbrace1+x \left(-\frac{4075}{4032}+\frac{67 \nu }{48}\right)+x^2
\left(-\frac{151877213}{67060224}-\frac{123815 \nu }{44352}+\frac{205 \nu ^2}{352}\right)+\pi x^{5/2} \left(-\frac{253 }{336}+\frac{253  \nu }{84}\right)\nonumber\\*
&\phantom{=.\frac{2}{7}\sqrt{5}\Bigg\lbrace}+x^3
\left(-\frac{4397711103307}{532580106240}+\left(\frac{700464542023}{13948526592}-\frac{205 \pi ^2}{96}\right) \nu +\frac{69527951 \nu ^2}{166053888}+\frac{1321981
\nu ^3}{5930496}\right)\Bigg\rbrace,\\
\mathcal{P}_{\text{spin}}^{(2,0)}&=\frac{2}{7}\sqrt{5}\Bigg\lbrace x^{3/2} \left(\frac{16 S_\ell}{3 M^2}+\frac{419 \delta  \Sigma _\ell}{160 M^2}\right)\nonumber\\
&\phantom{=.\frac{2}{7}\sqrt{5}\Bigg\lbrace}+x^2
\left(-\frac{128 \vec{S}\cdot\vec{S}+24 \vec{\Sigma}\cdot\vec{\Sigma} +256 S_\ell^2+75
\Sigma _\ell^2}{96 M^4}-\frac{4 \delta  \left(\vec{S}\cdot\vec{\Sigma} +2 S_\ell \Sigma _\ell\right)}{3 M^4}+\frac{4 \left(\vec{\Sigma}\cdot\vec{\Sigma} +2 \Sigma _\ell^2\right)\nu}{3 M^4}\right)\Bigg\rbrace,\\
\mathcal{P}^{(3,3)}&=-\frac{44 x^3 \delta  \nu }{27 \sqrt{35}},\\
\mathcal{P}_{\text{spin}}^{(3,3)}&=0,\\
\mathcal{P}^{(3,1)}&=\frac{223}{120\sqrt{21}}\Bigg\lbrace x^3\left(\frac{3872 \delta  \nu}{223}\right)\Bigg\rbrace,\\
\mathcal{P}_{\text{spin}}^{(3,1)}&=0,\\
\mathcal{P}^{(4,4)}&=-\frac{4}{3} i \sqrt{\frac{2}{35}} x^{5/2} \nu,\\
\mathcal{P}_{\text{spin}}^{(4,4)}&=0,\\
\mathcal{P}^{(4,1)}&=0,\\
\mathcal{P}_{\text{spin}}^{(4,1)}&=\frac{13}{56\sqrt{5}}\Bigg\lbrace x^{3/2} \left(-\frac{\left(S_n-i S_{\lambda }\right)}{M^2}-\frac{34 \delta  \left(\Sigma _n-i \Sigma _{\lambda }\right)}{39 M^2}\right)\nonumber\\
&\phantom{=.\frac{13}{56\sqrt{5}}\Bigg\lbrace}+x^2 \Bigg(\frac{10 \left(3 S_l \left(S_n-i S_{\lambda }\right)+\Sigma _l
	\left(\Sigma _n-i \Sigma _{\lambda }\right)\right)}{117 M^4}+\frac{5 \delta  \left(\left(S_n-i
	S_{\lambda }\right) \Sigma _l+S_l \left(\Sigma _n-i \Sigma _{\lambda }\right)\right)}{39 M^4}\nonumber\\
&\phantom{=.\frac{13}{56\sqrt{5}}\Bigg\lbrace+x^{2}\Bigg(}-\frac{10 \nu  \Sigma _l \left(\Sigma _n-i \Sigma _{\lambda }\right)}{39 M^4}\Bigg)\Bigg\rbrace,\\
\mathcal{P}^{(4,0)}&=\frac{1}{42}\Bigg\lbrace1+x \left(-\frac{180101}{29568}+\frac{27227 \nu }{1056}\right)\nonumber\\
&\phantom{=.\frac{1}{42}\Bigg\lbrace}+x^2
\left(\frac{2201411267}{158505984}-\frac{34829479 \nu }{432432}+\frac{844951 \nu ^2}{27456}\right)+x^{5/2} \left(-\frac{13565 \pi }{1232}+\frac{13565 \pi  \nu }{308}\right)\nonumber\\
&\phantom{=.\frac{1}{42}\Bigg\lbrace}+x^3
\left(\frac{15240463356751}{781117489152}+\left(-\frac{1029744557245}{27897053184}-\frac{205 \pi ^2}{96}\right) \nu -\frac{4174614175 \nu
	^2}{36900864}+\frac{221405645 \nu ^3}{11860992}\right)\Bigg\rbrace,\\
\mathcal{P}_{\text{spin}}^{(4,0)}&=\frac{1}{42}\Bigg\lbrace x^{3/2} \left(\frac{10 S_\ell}{M^2}+\frac{57 \delta  \Sigma _\ell}{8 M^2}\right)\nonumber\\
&\phantom{=.\frac{1}{42}\Bigg\lbrace}+x^2
\left(-\frac{64 \vec{S}\cdot\vec{S}+12 \vec{\Sigma}\cdot\vec{\Sigma} +128
	S_\ell^2+41 \Sigma _\ell^2}{48 M^4}-\frac{4 \delta  \left(\vec{S}\cdot\vec{\Sigma} +2 S_\ell \Sigma _\ell\right)}{3 M^4}+\frac{4 \left(\vec{\Sigma}\cdot\vec{\Sigma} +2 \Sigma _\ell^2\right)\nu}{3 M^4}\right)\Bigg\rbrace,\\
\mathcal{P}^{(5,5)}&=-\frac{36 x^3 \delta  \nu }{5 \sqrt{77}},\\
\mathcal{P}_{\text{spin}}^{(5,5)}&=0,\\
\mathcal{P}^{(5,3)}&=\frac{4 x^3 \delta  \nu }{27 \sqrt{385}},\\
\mathcal{P}_{\text{spin}}^{(5,3)}&=0,\\
\mathcal{P}^{(5,1)}&=\frac{2}{21}\sqrt\frac{2}{165}\Bigg\lbrace x^3\left(26\delta  \nu\right)\Bigg\rbrace,\\
\mathcal{P}_{\text{spin}}^{(5,1)}&=0,\\
\mathcal{P}^{(6,0)}&=-\frac{4195}{177408\sqrt{13}}x\Bigg\lbrace1-\frac{3612 \nu }{839}+x \left(-\frac{45661561}{6342840}+\frac{101414 \nu }{2517}-\frac{48118 \nu
	^2}{839}\right)+x^{3/2} \left(\frac{1248 \pi }{839}-\frac{4992 \pi  \nu }{839}\right)\nonumber\\
&\phantom{=.-\frac{4195}{177408\sqrt{13}}x\Bigg\lbrace}+x^2 \left(\frac{3012132889099}{144921208320}-\frac{27653500031 \nu }{191694720}+\frac{1317967427 \nu ^2}{4107744}-\frac{24793657 \nu
	^3}{342312}\right)\Bigg\rbrace,\\
\mathcal{P}_{\text{spin}}^{(6,0)}&=0,\\
\mathcal{P}^{(8,0)}&=\frac{75601}{8895744\sqrt{17}}x^2\Bigg\lbrace1-\frac{452070 \nu }{75601}+\frac{733320 \nu ^2}{75601}\nonumber\\
&\phantom{=.\frac{75601}{8895744\sqrt{17}}x^2}+x \left(-\frac{265361599}{33869248}+\frac{18177898147 \nu }{321757856}-\frac{722521125 \nu
	^2}{5745676}+\frac{261283995 \nu ^3}{2872838}\right)\Bigg\rbrace,\\
\mathcal{P}_{\text{spin}}^{(8,0)}&=0.
\end{align}
\end{subequations}
\end{widetext}

\section{PN STRAIN NEW MEMORY TERMS}
\label{sec:appendixb}

The post-Minkowski expansion of gravitational waves in radiative
coordinates yields the strain in terms of the radiative mass and current moments, $\mathrm{U}^{(\ell, m)}$ and $\mathrm{V}^{(\ell, m)}$
(see~\cite{Blanchet:2013haa}, but here following the conventions
of~\cite{Favata:2008yd}),
\begin{align}
  h^{(\ell, m)} = \frac{G}{\sqrt{2} R c^{\ell + 2}}
  \left( \mathrm{U}^{(\ell, m)} - \frac{i}{c} \mathrm{V}^{(\ell, m)}\right),
\end{align}
and when expressing the radiative moments in terms of the six types of
source moments, $\mathrm{U}^{(\ell, m)}$ is then even further decomposed in terms that are called
instantaneous, memory, tail, tail-of-tail, \ldots pieces:
\begin{align}
  \mathrm{U}^{(\ell, m)} =
  \mathrm{U}^{(\ell, m)}_{\text{inst}} +
  \mathrm{U}^{(\ell, m)}_{\text{tail}} +
  \mathrm{U}^{(\ell, m)}_{\text{tail-tail}} +
  \mathrm{U}^{(\ell, m)}_{\text{mem}} +
  \ldots
\end{align}
Therefore, we simply define
\begin{align}
  h_{\text{mem}}^{(\ell, m)} = \frac{G}{\sqrt{2} R c^{\ell + 2}} \mathrm{U}^{(\ell, m)}_{\text{mem}}.
\end{align}
Then the memory arises as (Eq.~(2.26) of~\cite{Favata:2008yd}, from
Eq.~(2.43c) of~\cite{Blanchet:1992br}, and changing from STF tensors
to spherical harmonics)
\begin{align}
  h_{\text{mem}}^{(\ell, m)}&= \frac{16\pi G}{R c^{4}} \sqrt{\frac{(\ell-2)!}{(\ell+2)!}} \times \nonumber\\
  &\phantom{=.\frac{16\pi G}{R c^{4}}}\int_{-\infty}^{u} du' \int d\Omega \frac{dE^{\text{GW}}}{dud\Omega} \overline{Y}^{(\ell, m)},
\end{align}
where
\begin{align}
  \frac{dE^{\text{GW}}}{du d\Omega} = \frac{R^{2}}{16\pi G} |\dot{h}|^{2}.
\end{align}
Consequently, we find
\begin{align}
  \label{eq:h_mem_in_terms_of_Psi_M}
  h_{\text{mem}}^{(\ell, m)} = \frac{4}{R c^{4}}
  \sqrt{\frac{(\ell-2)!}{(\ell+2)!}} \Psi^{(\ell, m)}_{\mathrm{M}},
\end{align}
or, in terms of the $\mathcal{P}^{(\ell, m)}$ decomposition of Eq.~\eqref{eq:Psi_M_decomp_calP},
\begin{align}
  \label{eq:h_mem_in_terms_of_calP}
h_{\text{mem}}^{(\ell,m)}=\frac{2 M\nu x}{R} 
\sqrt{\frac{(\ell-2)!}{(\ell+2)!}} \sqrt{4\pi} \mathcal{P}^{(\ell,m)}
e^{-im\psi}.
\end{align}
Another way to derive this expression is by writing
Eq.~\eqref{eq:reshear} with the strain, ignoring the contribution coming from
the Bondi mass aspect, and then moving the factor of $\eth^{2}$ to the
right-hand side.

Using Eq.~\eqref{eq:h_mem_in_terms_of_calP} and the
$\mathcal{P}^{(\ell,m)}$ expressions in
Eq.~\eqref{eq:PNMoreschisupermomentum} gives new oscillatory and
spin-dependent memory terms which have not been identified or have been missing from the PN
literature.

Finally we should point out the following interpretation of our BMS
frame fixing procedure, by studying
Eq.~\eqref{eq:h_mem_in_terms_of_Psi_M}.  Our procedure fixed the BMS
frame by matching $\Psi_{\mathrm{M}}$, mode by mode, to $\Psi_{\mathrm{M}}^{\text{PN}}$. Then, because of the simple
mode-wise proportionality
$h_{\text{mem}}^{(\ell,m)} \propto \Psi_{\text{M}}^{(\ell,m)}$, it turns out that our
frame-fixing is equivalent to matching $h_{\text{mem}}$ (except
that, being a spin-weight $-2$ field, $h_{\text{mem}}$ is missing the
$\ell=0,1$ information from $\Psi_{\text{M}}$).

\bibliography{main}

\end{document}